\documentclass[a4paper]{article}

\usepackage{graphicx}
\usepackage{epsfig}
\usepackage[fleqn]{amsmath}
\usepackage[psamsfonts]{amssymb}
\usepackage[varg]{txfonts}

\usepackage{lineno}

\title{{\bf Optimal Design Method of MIMO Antenna Directivities and Corresponding Current Distributions by Using Spherical Mode Expansion}}

\author{Maki Arai, Masashi Iwabuchi, Kei Sakaguchi, Kiyomichi Araki \\ 
	E-mail: arai@mobile.ee.titech.ac.jp}
\date{}

\newcommand{\argmax}{\mathop{\rm arg~max}\limits}
\newcommand{\mrm}{\mathrm}
\newcommand{\mbf}{\mathbf}
\newcommand{\mbb}{\mathbb}

\newcommand*\patchAmsMathEnvironmentForLineno[1]{%
  \expandafter\let\csname old#1\expandafter\endcsname\csname #1\endcsname
  \expandafter\let\csname oldend#1\expandafter\endcsname\csname end#1\endcsname
  \renewenvironment{#1}%
     {\linenomath\csname old#1\endcsname}%
     {\csname oldend#1\endcsname\endlinenomath}}%
\newcommand*\patchBothAmsMathEnvironmentsForLineno[1]{%
  \patchAmsMathEnvironmentForLineno{#1}%
  \patchAmsMathEnvironmentForLineno{#1*}}%
\AtBeginDocument{%
\patchBothAmsMathEnvironmentsForLineno{equation}%
\patchBothAmsMathEnvironmentsForLineno{align}%
\patchBothAmsMathEnvironmentsForLineno{flalign}%
\patchBothAmsMathEnvironmentsForLineno{alignat}%
\patchBothAmsMathEnvironmentsForLineno{gather}%
\patchBothAmsMathEnvironmentsForLineno{multline}%
}

\begin{document}
\maketitle
\section*{Summary}
This paper proposes a new methodology to design optimal antennas 
for MIMO (Multi-Input Multi-Output) communication systems 
by using spherical mode expansion. 
Given spatial channel properties of a MIMO channel, such as the angular profile at both sides, 
the optimal MIMO antennas should provide the largest channel capacity 
with a constraint of the limited implementation space (volume). 
In designing a conventional MIMO antenna, first the antenna structure (current distribution) is determined, 
second antenna directivity is calculated based on the current distribution, 
and thirdly MIMO channel capacity is calculated by using given angular profiles and obtained antenna directivity. 
This process is repeated by adjusting the antenna structure 
until the performance satisfies a predefined threshold. 
To the contrary, this paper solves the optimization problem analytically 
and finally gives near optimal antenna structure (current distribution) without any greedy search. 
In the proposed process, first the optimal directivity of MIMO antennas is derived 
by applying spherical mode expansion to the angular profiles, 
and second a far-near field conversion is applied on the derived optimal directivity 
to achieve near optimal current distributions on a limited surface. 
The effectiveness of the proposed design methodology is validated via numerical calculation of MIMO channel capacity as in the conventional design method 
while giving near optimal current distribution with constraint of an antenna structure derived from proposed methodology. 

{\bf keywords:}
MIMO, antenna directivity, antenna structure, current distribution, spherical mode expansion, capacity maximization.

\section{Introduction}
In the latest and future wireless communication systems, Multiple-Input Multiple-Output (MIMO) technology is important with respect to improving the system performance.
In MIMO systems, both the receiver and transmitter are constructed with multiple antenna elements
to increase the capacity in proportion to the number of antenna elements 
and to achieve better bit error rate performance by utilizing diversity and multiplexing gains
\cite{Telatar}\cite{Foschini}.

Designing the array antenna is difficult in terms of the size, polarization, 
mutual coupling, and spatial correlation between antenna elements \cite{Balanis}.
In order to reduce the antenna size, the antenna elements should be located 
in the smallest space possible.
This, however, degrades the capacity 
because the mutual coupling and spatial correlation become high when the distance between antenna elements is small \cite{Browne}.
To  decrease mutual coupling and spatial correlation, several solutions have been proposed.
For example, by using a capacitor or conductor connected between antenna elements, the effect of mutual coupling can be canceled \cite{Chen}.
Another example is using orthogonal polarization, such as horizontal and vertical polarization.
By using them, space diversity can be achieved with uncorrelated channels \cite{Kermoal}.
Conventionally, these problems are considered independently
and the optimal antenna to achieve the best performance has not been found yet.
Furthermore, the optimal radiation patterns have been derived by using an angular profile to improve diversity gain in the transmitter or receiver side \cite{Quist1}\cite{Quist2}. 
However, for the MIMO system, 
the propagation environments at the transmitter and receiver are not always independent.
Thus the directivity optimization of both transmitter and receiver sides are needed for the MIMO system.
To maximize the performance of MIMO systems, we should consider the problems comprehensively for designing the optimal antenna in both the transmitter and receiver sides.
To address this issue, we have proposed a new approach for antenna's design by using spherical mode expansion (SME) \cite{Hansen}\cite{Glazunov_ModeCorMatrix}.
SME has been used in many studies to analyze characteristics of antennas, circuits, propagation channels and so on \cite{Klemp_2005}-\cite{Miao}.
However, in these previous works, 
SME is used to just describe directivity of special type of antennas to evaluate the performance of MIMO systems.
On the other hand, 
we consider both the outer space expanded by SME (a propagation environment and antenna directivities) and the inner space (current distributions of antennas) 
by a far-near field conversion using SME
in order to maximize the performance of MIMO systems.

\begin{figure}[tb]
  \begin{center}
  \includegraphics[width=0.5\textwidth]{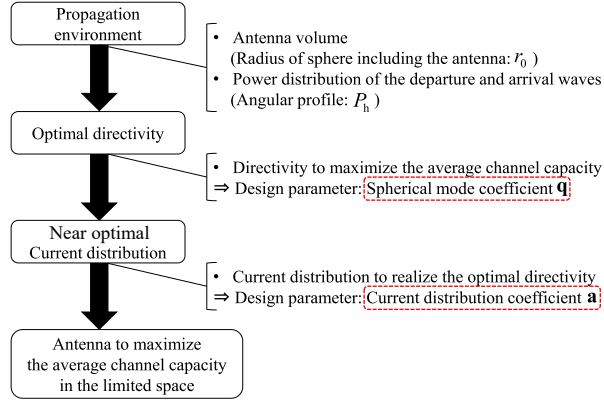}
  \end{center}
  \caption{Antenna design concept by using Spherical mode expansion.}
  \label{fig:AntennaDesignBySME}
\end{figure}

\begin{figure}[tb]
  \begin{center}
  \includegraphics[width=0.45\textwidth]{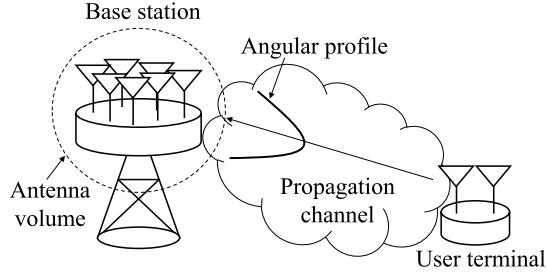}
  \end{center}
  \caption{MIMO system model.}
  \label{fig:SystemModel_MIMO}
\end{figure}

The concept of our proposed antenna design scheme is shown in Fig.\ \ref{fig:AntennaDesignBySME}.
First, the optimal directivity of each antenna element is derived from a power angular profile of departure and arrival waves.
In SME, an electrical and magnetic field is expanded by spherical wave functions 
and spherical mode coefficients (SMCs).
SMCs, which specify the antenna directivity, can be optimized from angular profiles of propagation environments 
drawn in Fig.\ \ref{fig:SystemModel_MIMO} 
if the antenna volume is given. 
Thus, designing antenna directivities is equivalent to calculating optimal SMCs to maximize the channel capacity.
Next, the current distribution to achieve the optimal directivity is calculated.
SMCs also determine the current distribution on the surface of the antenna volume, 
therefore the current distribution for the optimal directivity can be calculated 
by projecting it on the conductor surface to be implemented.
By using the above scheme, 
we can maximize channel capacity by matching antenna directivity to the propagation environment. 
In this paper, we will describe the detailed theory to derive optimal directivities and current distributions 
for MIMO antenna systems.
We will confirm the validity of the proposed method by comparing the channel capacity of the optimal directivity and the directivity recalculated from the current distributions to that of a conventional half-wave length dipole antenna array.

This paper is organized as follows.
In Sec. \ref{sec:SphericalModeExpansion}, spherical wave functions and SME of the antenna directivity are described. 
Section \ref{sec:DirectivityOptimization} describes the method of optimizing antenna directivities for the MIMO system.
In Sec. \ref{sec:CurrentDistribution}, the derivation method of near optimal current distribution with constraint of an antenna structure is described.
In Sec. \ref{sec:NumericalAnalysis}, the validity of the proposed method is confirmed by comparing the performance of the optimal directivity and the recalculated directivity.
In Sec. \ref{sec:Conclusion}, the conclusions are summarized.

\section{Spherical mode expansion}
\label{sec:SphericalModeExpansion}
How to express the antenna directivity by using SME is introduced in this section.
Additionally, truncation of the modes is described to define the number of effective SMCs in a limited antenna volume.

\subsection{Spherical wave function}
\begin{figure}[tb]
  \begin{center}
  \includegraphics[width=0.4\textwidth]{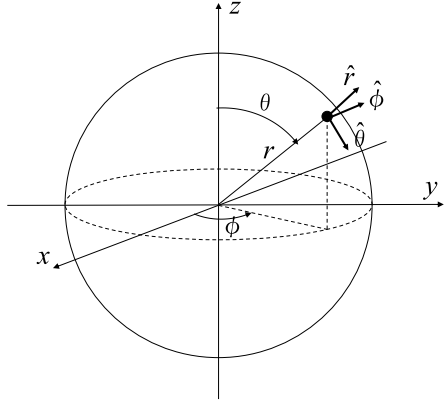}
  \end{center}
  \caption{Spherical coordinate.}
  \label{fig:SphericalCoordinate}
\end{figure}

Spherical wave functions are canonical solutions of the Helmholtz equation in spherical coordinates.
Since these functions have the orthogonality between different modes, 
linear analyses can be applied to any functions in the spherical coordinates as shown in Fig.\ \ref{fig:SphericalCoordinate}.
There are two groups of solutions, which are expressed as follows,
\begin{equation}
\begin{split}
	\vec{f}_{1mn}^{(c)}(r, \theta, \phi) &= \frac{1}{\sqrt{2\pi}} \frac{1}{\sqrt{n(n+1)}} \left( -\frac{m}{|m|} \right)^{m}\\
	& \Bigg\{ z_{n}^{(c)}(kr) \frac{im\bar{P}_{n}^{|m|}(\cos\theta)}{\sin\theta}e^{im\phi}\hat{\theta}\\
	& - z_{n}^{(c)}(kr) \frac{\mathrm{d}\bar{P}_{n}^{|m|}(\cos\theta)}{\mathrm{d}\theta}e^{im\phi}\hat{\phi} \Bigg\},\\
	\vec{f}_{2mn}^{(c)}(r, \theta, \phi) &= \frac{1}{\sqrt{2\pi}}\frac{1}{\sqrt{n(n+1)}} \Bigg( -\frac{m}{|m|} \Bigg)^{m}\\
	& \Bigg\{ \frac{n(n+1)}{kr}z_{n}^{(c)}(kr)\bar{P}_{n}^{|m|}(\cos\theta)e^{im\phi}\hat{r}\\
	& + \frac{1}{kr}\frac{\mathrm{d}}{\mathrm{d}(kr)}(krz_{n}^{(c)}(kr))
		\frac{\mathrm{d}\bar{P}_{n}^{|m|}(\cos\theta)}{\mathrm{d}\theta}e^{im\phi}\hat{\theta}\\
	& + \frac{1}{kr}\frac{\mathrm{d}}{\mathrm{d}(kr)}(krz_{n}^{(c)}(kr))
		\frac{im\bar{P}_{n}^{|m|}(\cos\theta)}{\sin\theta}e^{im\phi}\hat{\phi} \Bigg\},\label{eq:SWF}
\end{split}
\end{equation}
where $k$ is the wave number in free space and $i$ is the imaginary unit.
$\bar{P}^{|m|}_{n}(x)$ is the normalized associated Legendre function of $n(=$1,2,3,$\cdots)$-th degree 
and $m(=$$-n,-n+1\cdots 0\cdots n-1,n)$-th order, and $z^{(c)}_{n}(x)$ is the radial function 
shown in Table\ \ref{tbl:Radial} defined by $c=1,2,3,4$.
The radial function is specified by index $c$ and degree $n$.
Spherical wave functions $\vec{f}_{smn}^{(c)}$ have indices $s(=1,2)$, $m$, $n$ and $c$,
where index $s$ identifies the solution of the Helmholtz equation.
$s=1$ means a Transverse electric (TE) wave
and $s=2$ means a Transverse magnetic (TM) wave.
$\theta$ is the elevation angle and $\phi$ is the azimuth angle in spherical coordinate.
$\hat{r}$, $\hat{\theta}$ and $\hat{\phi}$ are unit vectors 
for corresponding directions of the spherical coordinate.
Table \ref{tbl:Radial} shows types of the index $c$.
For example, electric and magnetic fields are expressed by using the index $c=3$.
\begin{table}[htb]
	\begin{center}
	\caption{Radial function $z^{(c)}_{n}(x)$.}
	\label{tbl:Radial}
		\begin{tabular}{|c|l|l|}
		\hline
		Index $c$ & $z^{(c)}_{n}(x)$ & Function \\
		\hline
		$c=1$ & $j_n(x)$ & Spherical Bessel function \\
		 & & (a radial standing wave, finite at the origin)\\
		\hline
		$c=2$ & $n_n(x)$ & Spherical Neumann function \\
		 & & (a radial standing wave, infinite at the origin)\\
		\hline
		$c=3$ & $h^{(1)}_n(x)$ & Spherical Hankel function of the first kind \\
		 & & (a radial outgoing wave, infinite at the origin)\\
		\hline
		$c=4$ & $h^{(2)}_n(x)$ & Spherical Hankel function of the second kind \\
		 & & (a radial incoming wave, infinite at the origin)\\
		\hline
		\end{tabular}
	\end{center}
\end{table}

\subsection{Far-field pattern function}
Spherical wave function in far-field is called ``Far-field pattern function".
Far-field pattern function is represented by the spherical wave function of $c=3$ from the definition of Table \ref{tbl:Radial}.
\begin{equation}
	\vec{k}_{smn}(\theta, \phi) = \lim_{kr\to\infty} \left\{ \sqrt{4\pi} \frac{kr}{e^{-ikr}} \vec{f}_{smn}^{(3)}(r, \theta, \phi) \right\}. \label{eq:Far-field_def}
\end{equation}
And Eq.\ (\ref{eq:Far-field_def}) becomes,
\begin{eqnarray}
	\label{eq:k1}
	\vec{k}_{1mn}(\theta, \phi) &=& \sqrt{\frac{2}{n(n+1)}} \left( -\frac{m}{|m|} \right)^m e^{im\phi} (-i)^{n+1}\nonumber\\
	&&\Bigg\{ \frac{im\bar{P}^{|m|}_{n}(\cos\theta)}{\sin\theta}\hat{\theta}
		- \frac{\mathrm{d}\bar{P}^{|m|}_{n}(\cos\theta)}{\mathrm{d}\theta}\hat{\phi} \Bigg\}, \\
	\label{eq:k2}
	\vec{k}_{2mn}(\theta, \phi) &=& \sqrt{\frac{2}{n(n+1)}} \left( -\frac{m}{|m|} \right)^m e^{im\phi} (-i)^{n}\nonumber\\
	&&\Bigg\{ \frac{\mathrm{d}\bar{P}^{|m|}_{n}(\cos\theta)}{\mathrm{d}\theta}\hat{\theta}
		+ \frac{im\bar{P}^{|m|}_{n}(\cos\theta)}{\sin\theta}\hat{\phi} \Bigg\}. \label{eq:Far-field}
\end{eqnarray}
Since these functions contain the vertical and horizontal polarization expressed as 
terms of $\hat{\theta}$ and $\hat{\phi}$ respectively, 
SME enables to consider the two types of the polarization at the same time.

\subsection{Spherical mode expansion of antenna directivity}
The antenna directivity can be expressed by using far-field pattern functions as follows,
\begin{equation}
	\vec{g}(\theta,\phi) = \sum_{smn} q_{smn} \vec{k}_{smn}(\theta,\phi)
	\label{eq:SME_direct}
\end{equation}
$\vec{g}(\theta, \phi)$ is the antenna directivity including $\theta$ and $\phi$ polarization.
We call $\theta$ and $\phi$ polarization as vertical and horizontal polarization respectively.
By using SME, each directivity can be expressed as a superposition of far-field pattern functions.
Thus, we can design the directivities by considering only coefficients $q_{smn}$ corresponding to the mode of far-field pattern function, which is called ``spherical mode coefficient (SMC)".

\subsection{Truncation of modes}
For convenience, the indices of SME $s$, $m$, $n$ are replaced by a single index $j$ from now on.
Relationship between indices $s$, $m$, $n$ and the index $j$ is expressed as follows,
\begin{equation}
	j = 2(n^2+n-1+m)+s. \label{eq:Relationsmn}
\end{equation}
Since the maximum value of the index $m$ is determined by the index $n$, 
the number of index $j$ depends on the maximum value of index $n$.
If the volume of target is limited, 
the number of index $n$ can be truncated at $n=N$ defined from the radius of the volume \cite{Hansen}
because the function can be sampled and recalculated 
by using limited number of modes in the spherical coordinates 
like a sampling theorem.
The number of modes depends on the truncation index $N$ as follows, 
\begin{eqnarray}
	N &=& \lfloor kr_0  \rfloor, \label{eq:Truncation} \\
	J &=& 2N(N+2), \label{eq:NumOfMode}
\end{eqnarray}
where a symbol $\lfloor \cdot \rfloor$ means a floor function which indicates the largest integer smaller than or equal to $kr_0$.
By using the index $j$, Eq.\ (\ref{eq:SME_direct}) can be rewritten into a vector form as follows.
\begin{eqnarray}
\label{eq:SME_direct2}
	\vec{g}(\theta,\phi) &=& \sum_{j=1}^J q_{j} \vec{k}_{j}(\theta,\phi)
	= \mbf{q}^\mrm{T} \vec{\mbf{k}}(\theta,\phi) \\
	\mbf{q} &=& \left[ q_1, \cdots, q_J \right]^\mrm{T} \\
	\vec{\mbf{k}}(\theta,\phi) &=& \left[ \vec{k}_1(\theta,\phi), \cdots, \vec{k}_J(\theta,\phi) \right]^\mrm{T},
\end{eqnarray}
where $\mbf{q} \in \mbb{C}^{J}$ is a vector of SMCs and 
$\vec{\mbf{k}}(\theta,\phi) \in \mbb{C}^{J}$ is a vector of far-field pattern functions 
which are vector functions with $\theta$ and $\phi$ polarization components defined in Eq.\ (\ref{eq:k1}) and (\ref{eq:k2}).
Since each of far-field pattern function is unique and does not depend on the propagation environments, 
designing the optimal antenna directivity is equivalent to deriving the vector of SMCs.

\section{Optimization of MIMO antenna directivity}
\label{sec:DirectivityOptimization}
In this section, we shall derive the optimal antenna directivities of MIMO systems 
by using mathematical tools of SME shown in Sect.\ \ref{sec:SphericalModeExpansion}. 
The definition of optimization is to maximize the average channel capacity 
given a joint angular profile of the propagation channel.

\subsection{MIMO system model}
\begin{figure}[tb]
    \centering
    \includegraphics[width=0.5\textwidth]{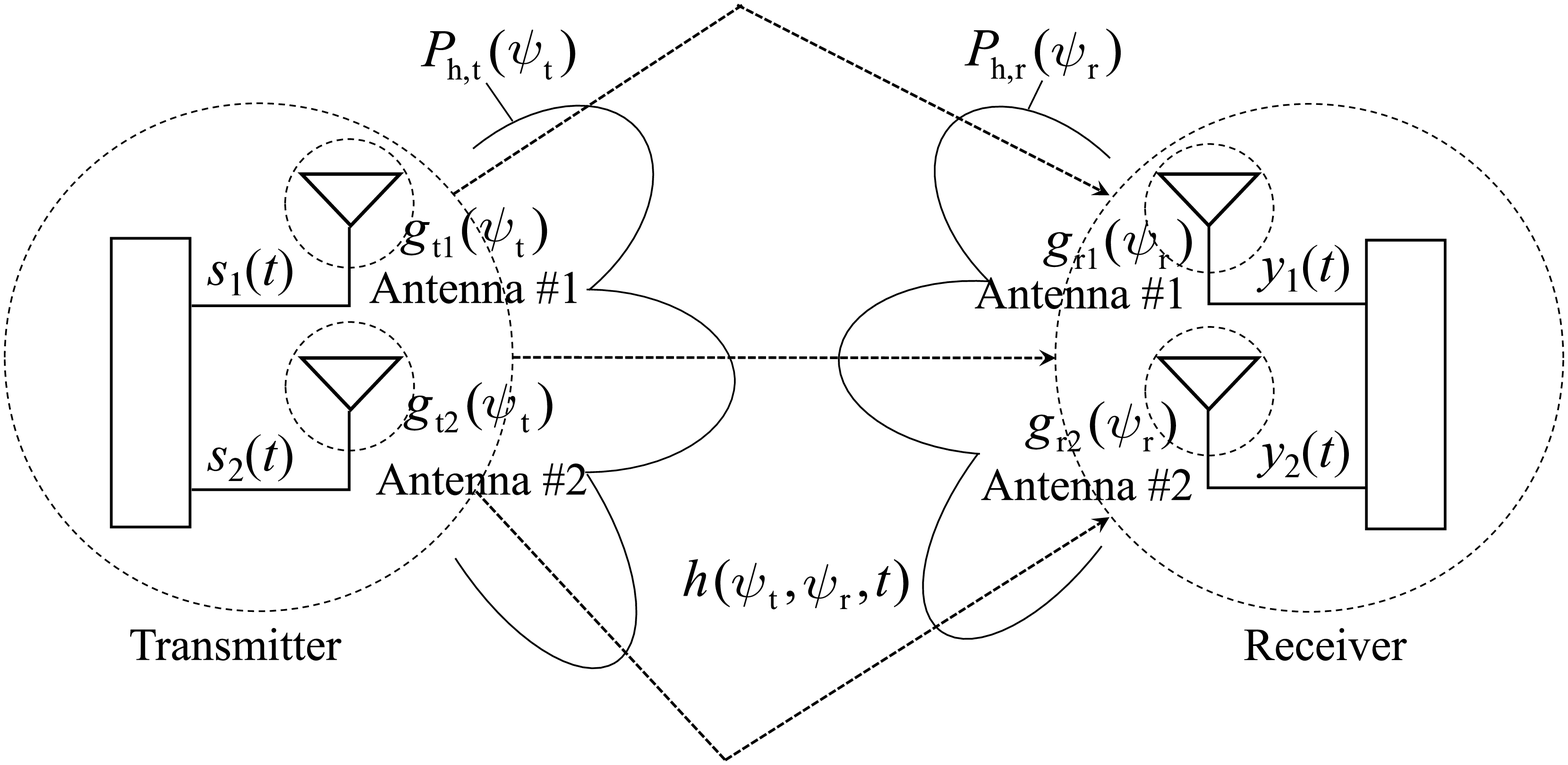}
    \caption{MIMO system model including antenna directivities ($N_{\mrm{t}}$ = $N_{\mrm{r}}$ = 2).}
    \label{fig:2x2MIMO_System}
\end{figure}

$N_{\mrm{t}} \times N_{\mrm{r}}$ MIMO system is shown in Fig.\ \ref{fig:2x2MIMO_System}.
The angle of departure and the angle of arrival are defined as 
$\psi_{\mrm{t}}=(\theta_{\mrm{t}}, \phi_{\mrm{t}})$ and $\psi_{\mrm{r}}=(\theta_{\mrm{r}}, \phi_{\mrm{r}})$.
The vector of transmit antenna directivities $\vec{\mbf{g}}_{\mrm{t}} \in \mbb{C}^{N_{\mrm{t}}}$ 
and that of receive antenna directivities $\vec{\mbf{g}}_{\mrm{r}} \in \mbb{C}^{N_{\mrm{r}}}$ are defined as follows, 
\begin{eqnarray}
\label{eq:Tx_Directivity}
    \vec{\mathbf{g}}_{\mathrm{t}}(\psi_{\mathrm{t}}) 
    = \left[\vec{g}_{\mathrm{t}1}(\psi_{\mathrm{t}}), \cdots, 
    \vec{g}_{{\mathrm{t}}N_{\mathrm{t}}}(\psi_{\mathrm{t}}) \right]^{\mathrm{T}}, \\
\label{eq:Rx_Directivity}
    \vec{\mathbf{g}}_{\mathrm{r}}(\psi_{\mathrm{r}}) 
    = \left[\vec{g}_{\mathrm{r}1}(\psi_{\mathrm{r}}), \cdots, 
    \vec{g}_{{\mathrm{r}}N_{\mathrm{r}}}(\psi_{\mathrm{r}}) \right]^{\mathrm{T}},
\end{eqnarray}
where $\vec{g}_{{\mrm{t}}n_{\mrm{t}}}(\psi_{\mrm{t}})$($n_{\mrm{t}}=1,\cdots,N_{\mrm{t}}$) is the $n_{\mrm{t}}$-th transmit antenna directivity 
and $\vec{g}_{{\mrm{r}}n_{\mrm{r}}}(\psi_{\mrm{r}})$($n_{\mrm{r}}=1,\cdots,N_{\mrm{r}}$) is the $n_{\mrm{r}}$-th receive antenna directivity 
which we would like to optimize in this paper.
These directivities are vector functions with $\theta$ and $\phi$ polarization components defined in Eq.\ (\ref{eq:k1}) and (\ref{eq:k2}).

The received signal vector $\mbf{y}(t)  \in \mbb{C}^{N_{\mrm{r}}}$ is defined as
\begin{eqnarray}
\label{eq:SignalModel_MIMO}
    \mbf{y}(t) &=& \int_{\psi_{\mrm{r}}} \int_{\psi_{\mrm{t}}}
    \vec{\mbf{g}}_{\mrm{r}}(\psi_{\mrm{r}}) \cdot \vec{\vec{h}}(\psi_{\mrm{t}},\psi_{\mrm{r}},t) 
    \cdot \vec{\mbf{g}}^{\mrm{T}}_{\mrm{t}}(\psi_{\mrm{t}}) 
    \mrm{d} \psi_{\mrm{t}} \mrm{d} \psi_{\mrm{r}} \mbf{s}(t)  \nonumber \\ 
    && + \mbf{n}(t) \nonumber \\
    &=& \mbf{H}(t) \mbf{s}(t) + \mbf{n}(t), \\
\label{eq:ChannelMatrix}
    \mbf{H}(t) &=& \int_{\psi_{\mrm{r}}} \int_{\psi_{\mrm{t}}}
    \vec{\mbf{g}}_{\mrm{r}}(\psi_{\mrm{r}}) \cdot \vec{\vec{h}}(\psi_{\mrm{t}},\psi_{\mrm{r}},t) 
    \cdot \vec{\mbf{g}}^{\mrm{T}}_{\mrm{t}}(\psi_{\mrm{t}})
    \mrm{d}\psi_{\mrm{t}} \mrm{d} \psi_{\mrm{r}},
\end{eqnarray}
where $\mbf{H}(t) \in \mbb{C}^{N_{\mrm{r}} \times N_{\mrm{t}}}$ is a channel matrix including antenna directivities, 
$\vec{\vec{h}}(\psi_{\mrm{t}},\psi_{\mrm{r}},t)$ is a channel response including polarization components of the transmit and receive antennas,
where the symbol of the double arrow means the vector function determined four polarization components such as $\theta$ and $\phi$ polarizations in the transmitter and $\theta$ and $\phi$ polarizations in the receiver,
and $\mbf{s}(t)  \in \mbb{C}^{N_{\mrm{t}}}$ is the transmit signal vector.
The noise vector $\mbf{n}(t) \in \mbb{C}^{N_{\mrm{r}}}$ is expressed as 
$\mbf{n}(t) = \left[n_1(t), \cdots, n_{N_{\mrm{r}}}(t)\right]^{\mrm{T}}$ and $\mrm{E}[ \mbf{n}(t)\mbf{n}^\mrm{H}(t) ] = P_{\mrm{n}} \mbf{I}_{N_{\mrm{r}}}$ is assumed, 
where $\mbf{I}_{N_{\mrm{r}}}$ is a ${N_{\mrm{r}}} \times {N_{\mrm{r}}}$ unit matrix.
The total transmit power is $P=\mrm{E}[ \mbf{s}^\mrm{H}(t) \mbf{s}(t) ]$.
It is noted that, in the MIMO system, 
the angular profile in the receiver side depends on the antenna directivity on the transmitter side and vice versa. 
These angular profiles are calculated using a joint angular profile 
$\vec{\vec{P}}_{\mrm{h}} (\psi_{\mrm{t}},\psi_{\mrm{r}}) = \mrm{E} \bigg[\left| \vec{\vec{h}}(\psi_{\mrm{t}},\psi_{\mrm{r}},t) \right| ^2 \bigg]$ 
as follows.
\begin{eqnarray}
\label{eq:AngularProfile_Rx}
    \vec{P}_{\mrm{h,r}} (\psi_{\mrm{r}}) 
    &=& \mrm{E} \bigg[ \int_{\psi_{\mrm{t}}}
	\left| \vec{\vec{h}}(\psi_{\mrm{t}},\psi_{\mrm{r}},t) \right| ^2
	\vec{\mbf{g}}^{\mrm{T}}_{\mrm{t}}(\psi_{\mrm{t}}) \vec{\mbf{g}}_{\mrm{t}}(\psi_{\mrm{t}})
    \mrm{d}\psi_{\mrm{t}} \bigg] \nonumber \\
	&=& \int_{\psi_{\mrm{t}}}
     \vec{\vec{P}}_{\mrm{h}} (\psi_{\mrm{t}},\psi_{\mrm{r}})
	\vec{\mbf{g}}^{\mrm{T}}_{\mrm{t}}(\psi_{\mrm{t}}) \vec{\mbf{g}}_{\mrm{t}}(\psi_{\mrm{t}})
    \mrm{d}\psi_{\mrm{t}}, \\
\label{eq:AngularProfile_Tx}
    \vec{P}_{\mrm{h,t}} (\psi_{\mrm{t}}) 
    &=& \mrm{E} \bigg[ \int_{\psi_{\mrm{r}}}
	\left| \vec{\vec{h}}(\psi_{\mrm{t}},\psi_{\mrm{r}},t) \right| ^2
	\vec{\mbf{g}}^{\mrm{T}}_{\mrm{r}}(\psi_{\mrm{r}}) \vec{\mbf{g}}_{\mrm{r}}(\psi_{\mrm{r}})
    \mrm{d}\psi_{\mrm{r}} \bigg] \nonumber \\
	&=& \int_{\psi_{\mrm{r}}}
 \vec{\vec{P}}_{\mrm{h}} (\psi_{\mrm{t}},\psi_{\mrm{r}}) 
	\vec{\mbf{g}}^{\mrm{T}}_{\mrm{r}}(\psi_{\mrm{r}}) \vec{\mbf{g}}_{\mrm{r}}(\psi_{\mrm{r}})
    \mrm{d}\psi_{\mrm{r}}.
\end{eqnarray}
Since the angular profile in the receiver side is determined 
by the channel response and the directivities of transmit antennas, 
it is defined by using $\vec{\mbf{g}}_{\mrm{t}}(\psi_{\mrm{t}})$ 
and not depends on $\vec{\mbf{g}}_{\mrm{r}}(\psi_{\mrm{r}})$ in Eq.\ (\ref{eq:AngularProfile_Rx}). 
The angular profile in the transmitter side is defined and not depends on $\vec{\mbf{g}}_{\mrm{t}}(\psi_{\mrm{t}})$ in the similar way as shown in Eq.\ (\ref{eq:AngularProfile_Tx}).

\subsection{Correlation matrix and average channel capacity}
In this paper, we derive the optimal antenna directivity to maximize the average channel capacity.
The channel capacity of the $N_{\mrm{t}} \times N_{\mrm{r}}$ MIMO system is derived in \cite{Telatar},
and expressed as follows.
\begin{eqnarray}
\label{eq:Ave_Capacity}
    \bar{C} = \mrm{E}\left[ \log_2 \det\left( \mbf{I}_{N_{\mrm{r}}}+ \gamma_0 \mbf{H}(t)\mbf{H}^{\mrm{H}}(t)  \right) \right],
\end{eqnarray}
where $\gamma_0=P_{\mrm{s}}/P_{\mrm{n}}$ is the ratio of the transmit power and the noise power, 
$P_{\mrm{s}}=P/N_\mrm{min}$ is the transmit power per each stream and $N_\mrm{min}=\min \{N_{\mrm{t}}, N_{\mrm{r}} \}$ is the number of streams. 
It is noted that the power is divided equally for all streams in this paper for simple analysis.
Channel correlation matrix in the receiver side $\bar{\mbf{R}}_{\mrm{c,r}} \in \mbb{C}^{N_{\mrm{r}} \times N_{\mrm{r}}}$ 
is defined by using the channel matrix as follows.
\begin{eqnarray}
\label{eq:CorMatrix_Rx}
    \bar{\mbf{R}}_{\mrm{c,r}} &=& \mrm{E}\left[ \mbf{H}(t)\mbf{H}^\mrm{H}(t) \right].
\end{eqnarray}
The infimum of average channel capacity is related to the channel correlation matrices 
for a large signal-to-noise ratio (SNR)
as shown in \cite{Zhang_CapacityBound}.
In this case, maximizing the average channel capacity is equivalent to maximize the determinant of the correlation matrices including antenna directivities as follows.
\begin{eqnarray}
\label{eq:Capacity_and_CorMatrix}
	\max \bar{C} &\Leftrightarrow& \max \mrm{Inf} \ \bar{C} \nonumber \\
	&\Leftrightarrow& \max \log_2 \left( \det \bar{\mbf{R}}_{\mrm{c,r}} \right) \nonumber \\
	&\Leftrightarrow& \max \det \bar{\mbf{R}}_{\mrm{c,r}}.
\end{eqnarray}

On the other hand, the channel correlation matrix of the receiver can be rewritten by using SME as follows.
\begin{eqnarray}
\label{eq:CorMatrix_Rx2}
    \bar{\mbf{R}}_{\mrm{c,r}}
    &=& \mrm{E} \Bigl[ \int_{\psi_{\mrm{r}}} \int_{\psi_{\mrm{t}}}
    \Bigl( \vec{\mbf{g}}_{\mrm{r}}(\psi_{\mrm{r}}) \cdot \vec{\vec{h}}(\psi_{\mrm{t}},\psi_{\mrm{r}},t) 
    \cdot \vec{\mbf{g}}^{\mrm{T}}_{\mrm{t}}(\psi_{\mrm{t}}) \Bigr) \nonumber \\
    && \ \ \ \ \ \ \ \ \ \ \Bigl(
    \vec{\mbf{g}}^{*}_{\mrm{t}}(\psi_{\mrm{t}}) \cdot \vec{\vec{h}}^{*}(\psi_{\mrm{t}},\psi_{\mrm{r}},t)
    \cdot \vec{\mbf{g}}^{\mrm{H}}_{\mrm{r}}(\psi_{\mrm{r}}) \Bigr)
    \mrm{d}\psi_{\mrm{t}} \mrm{d} \psi_{\mrm{r}} 
    \Bigr] \nonumber \\
    &=& \int_{\psi_{\mrm{r}}} \int_{\psi_{\mrm{t}}}
    \vec{\mbf{g}}_{\mrm{r}}(\psi_{\mrm{r}}) \cdot 
    \Bigl( \vec{\mbf{g}}^{\mrm{T}}_{\mrm{t}}(\psi_{\mrm{t}}) 
    \cdot \vec{\vec{P}}_{\mrm{h}}(\psi_{\mrm{t}},\psi_{\mrm{r}}) \cdot
    \vec{\mbf{g}}^{*}_{\mrm{t}}(\psi_{\mrm{t}}) \Bigr) \nonumber \\
    && \ \ \ \ \ \ \ \ \ \ 
    \cdot \vec{\mbf{g}}^{\mrm{H}}_{\mrm{r}}(\psi_{\mrm{r}}) 
    \mrm{d}\psi_{\mrm{t}} \mrm{d} \psi_{\mrm{r}} \nonumber \\
    &=& \int_{\psi_{\mrm{r}}} 
    \vec{\mbf{g}}_{\mrm{r}}(\psi_{\mrm{r}}) \cdot 
    \vec{P}_{\mrm{h,r}} (\psi_{\mrm{r}})  
    \cdot \vec{\mbf{g}}^{\mrm{H}}_{\mrm{r}}(\psi_{\mrm{r}}) 
    \mrm{d} \psi_{\mrm{r}},
\end{eqnarray}
where the directivities of the receiver $\vec{\mbf{g}}_{\mrm{r}}(\psi_{\mrm{r}})$ 
can be expressed by using a matrix of SMCs $\mathbf{Q}_{\mrm{r}} \in \mbb{C}^{J \times N_\mrm{r}}$ and 
the vector of the far-field pattern functions $\vec{\mbf{k}}_{\mrm{r}}(\psi_{\mrm{r}})  \in \mbb{C}^{J}$ as
\begin{eqnarray}
\label{eq:Rx_Directivity_SME}
    \vec{\mbf{g}}_{\mrm{r}}(\psi_{\mrm{r}}) &=& \mbf{Q}^{\mrm{T}}_{\mrm{r}} \vec{\mbf{k}}_{\mrm{r}}(\psi_{\mrm{r}}), \\
\label{eq:Rx_Q}
    \mbf{Q}_{\mrm{r}} &=& \left[ \mbf{q}_\mrm{r1}, \cdots, \mbf{q}_{\mrm{r}{N_{\mrm{r}}}} \right], \\
    \vec{\mbf{k}}_{\mrm{r}}(\psi_{\mrm{r}}) &=& \left[ \vec{k}_1(\psi_{\mrm{r}}), \cdots, \vec{k}_J(\psi_{\mrm{r}}) \right],
\end{eqnarray}
By substituting Eq.\ (\ref{eq:Rx_Directivity_SME}) into (\ref{eq:CorMatrix_Rx2}), 
the channel correlation matrix can be expressed by quadratic form as, 
\begin{eqnarray}
\label{eq:R_cr}
	\bar{\mbf{R}}_{\mrm{c,r}} &=& \mbf{Q}^{\mrm{T}}_{\mrm{r}} \int_{\psi_{\mrm{r}}} 
    \vec{\mbf{k}}_{\mrm{r}}(\psi_{\mrm{r}}) \cdot 
    \vec{P}_{\mrm{h,r}} (\psi_{\mrm{r}})  
    \cdot \vec{\mbf{k}}^{\mrm{H}}_{\mrm{r}}(\psi_{\mrm{r}})
    \mrm{d} \psi_{\mrm{r}} \mbf{Q}^{\mrm{*}}_{\mrm{r}} \nonumber \\
	&=& \mbf{Q}^{\mrm{T}}_{\mrm{r}} \mbf{R}_{\mrm{r}} \mbf{Q}^{*}_{\mrm{r}}.
\end{eqnarray}
$\mbf{R}_{\mrm{r}}  \in \mbb{C}^{J \times J}$ is called spherical mode correlation matrix at the receiver side
which is calculated by a combination of the angular profile $\vec{P}_{\mrm{h,r}} (\psi_{\mrm{r}})$ and the far-field pattern functions $\vec{\mbf{k}}_{\mrm{r}}(\psi_{\mrm{r}})$.
Therefore, the antenna directivities of the receiver can be optimally designed 
with the knowledge of $\mbf{R}_{\mrm{r}}$. 
Similarly, the antenna directivities of the transmitter can be optimized 
with the knowledge of spherical mode correlation matrix at the transmitter side $\mbf{R}_{\mrm{t}}$.

\subsection{Optimal spherical mode coefficients and directivity}
\label{sec:OptimalSMC}
As a final step, 
the optimal antenna directivities to maximize the average channel capacity is explained. 
In this subsection, we'll concentrate on expressions in the receiver side, 
however, the directivities of the transmitter can be optimized similarly 
based on the given directivities at the receiver.
From Eq.\ (\ref{eq:Capacity_and_CorMatrix}), 
maximizing the average channel capacity is equal to maximizing the determinant of the channel correlation matrix.
By substituting Eq.\ (\ref{eq:R_cr}) into (\ref{eq:Capacity_and_CorMatrix}), 
the determinant of the channel correlation matrix can be maximized 
by controlling the SMC matrix $\mathbf{Q}_{\mrm{r}}$. 
Since the channel correlation matrix is semi-positive definite matrix, 
it can be transformed by the eigenvalue decomposition using the SMC matrix 
as shown in Eq.\ (\ref{eq:Rx_Q}) and (\ref{eq:R_cr}), 
where $|\mbf{q}_1|^2\!=\!\cdots\!=\!|\mbf{q}_{N_\mrm{r}}|^2\!=\!1$.
The maximum determinant of the channel correlation matrix is expressed 
by Hadamard inequality \cite{Beckenbach} as follows. 
\begin{eqnarray}
	\max \det \bar{\mbf{R}}_{\mrm{c,r}} 
	= \max \det (\mbf{Q}^{\mrm{T}}_{\mrm{r}} \mbf{R}_{\mrm{r}} \mbf{Q}^{*}_{\mrm{r}}) 
	\leq \prod_{j=1}^{N_{\mrm{r}}} (\mbf{q}^{\mrm{T}}_{\mrm{r}j} \mbf{R}_{\mrm{r}} \mbf{q}^{*}_{\mrm{r}j}). 
\end{eqnarray}
The equality is achieved when $\mbf{q}^{\mrm{T}}_{\mrm{r}i} \mbf{R}_{\mrm{r}} \mbf{q}^{*}_{\mrm{r}j}\!=\!0 \ (i \! \neq \! j)$ is satisfied. 
Thus, the vectors to maximize the determinant of the channel correlation matrix are derived by the eigen vectors from the first to the $N_\mrm{r}$-th order of $\mbf{R}_{\mrm{r}}$. 
\begin{eqnarray}
	[\mbf{u}_{\mrm{r}1}, \cdots, \mbf{u}_{\mrm{r}N_\mrm{r}}]
	&=& \argmax_{\mbf{q}_1^*, \cdots, \mbf{q}_{N_\mrm{r}}^*}
	\mrm{det} (\mbf{Q}^{\mrm{T}}_{\mrm{r}} \mbf{R}_{\mrm{r}} \mbf{Q}^{*}_{\mrm{r}}), 
\end{eqnarray}
Therefore, the maximum determinant of the channel correlation matrix is derived as
\begin{eqnarray}
\label{eq:Maximize_Det_Rx}
	\max \det \bar{\mbf{R}}_{\mrm{c,r}} 
	&=& \max \det (\mbf{Q}^{\mrm{T}}_{\mrm{r}} \mbf{R}_{\mrm{r}} \mbf{Q}^{*}_{\mrm{r}}) \nonumber \\
	&=& \prod_{j=1}^{N_{\mrm{r}}} (\mbf{u}^{\mrm{H}}_{\mrm{r}j} \mbf{R}_{\mrm{r}} \mbf{u}_{\mrm{r}j}) 
	= \prod_{j=1}^{N_{\mrm{r}}} \lambda_{\mrm{r}j}.
\end{eqnarray}
$\lambda_j (j=1,\cdots,J)$ is an eigenvalue of the spherical mode correlation matrix 
$\mbf{R}_{\mrm{r}}$ that can be calculated as
\begin{eqnarray}
\label{eq: R_r}
    \mbf{R}_{\mrm{r}} &=& \mbf{U}_{\mrm{r}} \mbf{\Lambda}_{\mrm{r}} \mbf{U}^{\mrm{H}}_{\mrm{r}},  \\
    \mbf{\Lambda}_{\mrm{r}} &=& \mrm{diag} \left[ \lambda_{\mrm{r}1}, \cdots, \lambda_{\mrm{r}J} \right],  \\
    &&(\lambda_{\mrm{r}j} \geq \lambda_{\mrm{r}(j+1)} \geq 0, \ \mrm{for}\  j=1, \cdots, J-1) \nonumber \\
    \mbf{U}_{\mrm{r}} &=& \left[ \mbf{u}_{\mrm{r}1}, \cdots, \mbf{u}_{\mrm{r}J} \right],
\end{eqnarray}
where $\mbf{U}_{\mrm{r}} \in \mbb{C}^{J \times J}$ is an eigen matrix.
Therefore, the optimal spherical coefficients $\mbf{Q}_{\mrm{r,opt}} \in \mbb{C}^{J \times N_\mrm{r}}$ can be calculated as a set of eigen vectors corresponding to 
the $N_\mrm{r}$ largest eigenvalues as
\begin{eqnarray}
\label{eq:Q_ropt}
    \mbf{Q}_{\mrm{r,opt}} = \left[ \mbf{u}^{*}_{\mrm{r}1}, \cdots, \mbf{u}^{*}_{\mrm{r}N_{\mrm{r}}} \right].
\end{eqnarray}
Finally the optimal antenna directivities are derived by using the optimal spherical coefficients
$\mbf{Q}_{\mrm{r,opt}}$ as follows.
\begin{eqnarray}
\label{eq:Optimal_gr}
    \vec{\mbf{g}}_{\mrm{r}}(\psi_{\mrm{r}}) 
    &=& \mbf{Q}^{\mrm{T}}_{\mrm{r,opt}} \vec{\mbf{k}}_{\mrm{r}}(\psi_{\mrm{r}}) \nonumber \\
    &=& \left[ \mbf{u}^{\mrm{H}}_{\mrm{r}1} \vec{\mbf{k}}_{\mrm{r}}(\psi_{\mrm{r}}), \cdots, 
    \mbf{u}^{\mrm{H}}_{\mrm{r}N_{\mrm{r}}} \vec{\mbf{k}}_{\mrm{r}}(\psi_{\mrm{r}}) \right]^{\mrm{T}}.
\end{eqnarray}

\subsection{Optimization at both transmitter and receiver sides}
\begin{figure}[tb]
    \centering
    \includegraphics[width=0.45\textwidth]{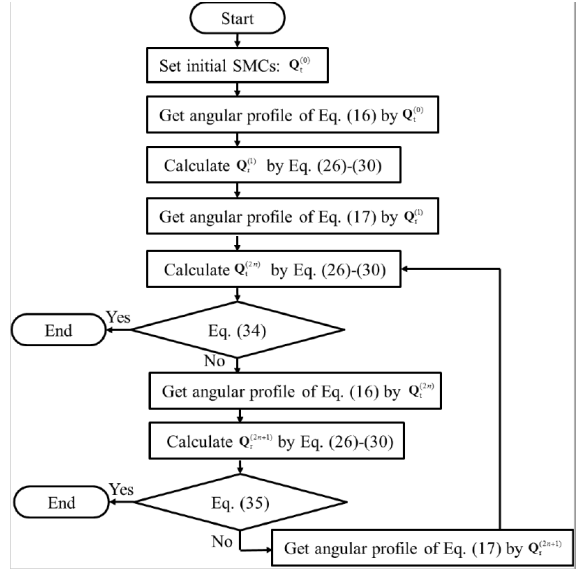}
    \caption{Sequential optimization at transmitter and receiver.}
    \label{fig:Sequential}
\end{figure}

In the case of optimization at either receiver or transmitter side, 
the optimal directivity is derived by the method in subsection \ref{sec:OptimalSMC}.
On the other hand, when the optimization at both receiver and transmitter sides is needed, 
one of solutions is a sequential optimization as shown in Fig.\ \ref{fig:Sequential}.
In this manuscript, we assume that the joint angular profile can be estimated ideally. 
Then, Fig.\ \ref{fig:Sequential} shows an algorithm to derive optimal directivities 
at the transmitter and receiver in an iterative manner by using the given joint angular profile. 
Therefore, no iterative power angular profile estimation is needed.

First, initial SMCs $\mbf{Q}_{\mrm{t}}^{(0)}$ are defined at the transmitter. 
Next, the SMCs of the receiver side $\mbf{Q}_{\mrm{r}}^{(1)}$ can be calculated by Eq.\ (\ref{eq: R_r}) and (\ref{eq:Q_ropt}) with the knowledge of angular profile containing $\mbf{Q}_{\mrm{t}}^{(0)}$.
In the same way, the SMCs of the transmitter and receiver sides, $\mbf{Q}_{\mrm{t}}^{(2n)}$ and $\mbf{Q}_{\mrm{r}}^{(2n+1)}$, can be calculated with the knowledge of angular profile containing $\mbf{Q}_{\mrm{r}}^{(2n-1)}$ and $\mbf{Q}_{\mrm{t}}^{(2n)}$ alternately.
These calculations should be repeated until the value of objective function converges.
The convergence conditions at transmitter and receiver are indicated respectively as follows.
\begin{eqnarray}
	\left|\det \bar{\mbf{R}}_{\mrm{c,t}}^{(2n)} - \det \bar{\mbf{R}}_{\mrm{c,r}}^{(2n-1)} \right| < \epsilon \\
	\left|\det \bar{\mbf{R}}_{\mrm{c,r}}^{(2n+1)} - \det \bar{\mbf{R}}_{\mrm{c,t}}^{(2n)} \right| < \epsilon,
\end{eqnarray}
where $\epsilon$ is a capable difference.
Consequently, SMCs at both transmitter and receiver sides are calculated by the sequantial optimization 
when the objective function converges as 
\begin{eqnarray}
	\lim_{itr \rightarrow \infty} \left|\det \bar{\mbf{R}}_{\mrm{c,t}}^{(itr+1)} - \det \bar{\mbf{R}}_{\mrm{c,r}}^{(itr)} \right|  = 0.
\end{eqnarray}

\section{Derivation of current distribution for optimal antenna directivity}
\label{sec:CurrentDistribution}
In this section, we shall derive the near optimal current distribution with constraint of an antenna structure
by using the far-near field conversion of the SME. 
For that purpose, a new matrix equation 
between coefficients of current distribution on an implementation surface 
and designed spherical mode coefficients in far-field pattern is developed.
Since the same procedure is used at the transmitter and receiver sides, 
we shall discuss the derivation of the near optimal current distribution generally 
by using SMCs' vector of the $n$-th antenna directivity $\mbf{q}_n$.

\subsection{Orthogonal basis functions and current distribution coefficients}
\begin{figure}[tb]
    \centering
    \includegraphics[width=0.3\textwidth]{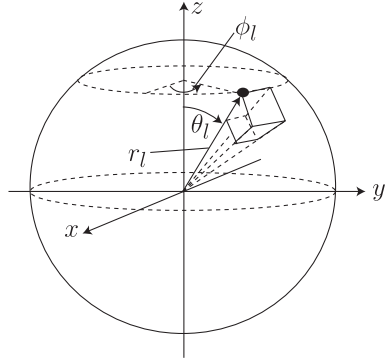}
    \caption{Minute region indicated by the index $l$.}
    \label{fig:GalerkinMethod}
\end{figure}

The current distribution occurs on the implementation surface of antenna included in the antenna volume with the radius $r_0$.
The current distribution has $r,\theta,\phi$ components in the spherical coordinate 
and expressed as the summation of the current on the minute region
when the implementation surface is divided into $L$ minute regions.
The current on the minute region is defined as $\vec{a}_{l} \vec{b}_{l} (r,\theta,\phi)$, 
where $\vec{a}_{l}$ is a current distribution coefficient
and $\vec{b}_{l} (r,\theta,\phi)$ is an orthogonal basis function defined as 
\begin{eqnarray}
\label{eq:BasisFunc}
    \vec{b}_l (r,\theta,\phi) 
    &=& b_{l}(r)\hat{r} + b_{l}(\theta)\hat{\theta} + b_{l}(\phi)\hat{\phi}, \\
\label{eq:BasisFunc2}
	\int_{V_0} b_{l} (u) b_{l'}(u) \mrm{d} V_0
    &=& \delta_{ll'}, \ \ (\mrm{for} \ \ u=r,\theta,\phi)
\end{eqnarray}
where $\delta_{ll'}$ indicates Kronecker's delta
and $V_0$ is the antenna volume and the index $l$ ($l=1,\cdots,L$) indicates the $l$-th minute region in the antenna volume as shown in Fig.\ \ref{fig:GalerkinMethod}.
Therefore, the current distribution on the implementation surface can be expanded 
by using orthogonal basis functions and current distribution coefficients as 
\begin{eqnarray} 
\label{eq:Current_Galerkin}
    \vec{J}(r,\theta,\phi) 
    &=& \sum_{l=1}^{L} \vec{a}_{l} \vec{b}_{l} (r,\theta,\phi) \nonumber \\
    &=& \sum_{l=1}^{L} \Big(
    a_{l}^{r} b_{l}^{r} (r,\theta,\phi) \hat{r} \nonumber \\
    && + a_{l}^{\theta} b_{l}^{\theta} (r,\theta,\phi) \hat{\theta}
    + a_{l}^{\phi} b_{l}^{\phi} (r,\theta,\phi) \hat{\phi} \Big).
\end{eqnarray}

\subsection{Current distribution for optimal antenna directivity}
From \cite{Hansen}, SMCs are derived from spherical wave functions of $c=1$ and a current distribution as
\begin{eqnarray}
\label{eq:SMC_from_Current}
    q_{smn} = (-1)^{m+1} \!\! \int_{V_0} \frac{k}{\sqrt[]{\mathstrut \eta}}
    \vec{f}^{(1)}_{s,-m,n}(r,\theta,\phi) \cdot \vec{J}(r,\theta,\phi) \mrm{d}V_0, 
\end{eqnarray}
where $\eta$ is a characteristic admittance and $\vec{J}(r,\theta,\phi)$ is 
the current distribution in Eq.\ (\ref{eq:Current_Galerkin}).
From Eqs.\ (\ref{eq:SMC_from_Current}) and (\ref{eq:Current_Galerkin}), the SMC can be expressed as 
\begin{eqnarray}
\label{eq:qj}
    q_j &=& (-1)^{m+1}  \!\! \int_{V_0}  \! \frac{k}{\sqrt[]{\mathstrut \eta}} 
    \vec{f}^{(1)}_{s,-m,n}(r,\theta,\phi) 
    \left( \sum_{l=1}^{L} \vec{a}_{l} \vec{b}_{l} (r,\theta,\phi) \right)
    \mrm{d}V_0 \nonumber \\
    &=&  \sum_{l=1}^{L} (-1)^{m+1} \frac{k}{\sqrt[]{\mathstrut \eta}}
    \biggl( 
    a_{l}^{r} \int_{V_0} f^{(1)r}_{s,-m,n}(r,\theta,\phi) b_{l} (r) \mrm{d}V_0 \nonumber \\
    && + a_{l}^{\theta} \int_{V_0} f^{(1)\theta}_{s,-m,n}(r,\theta,\phi) b_{l} (\theta) \mrm{d}V_0 \nonumber \\
    && + a_{l}^{\phi} \int_{V_0} f^{(1)\phi}_{s,-m,n}(r,\theta,\phi) b_{l} (\phi) \mrm{d}V_0
    \biggr),
\end{eqnarray}
where $f^{{(1)}r}_{s,-m,n}(r,\theta,\phi)$, $f^{{(1)}\theta}_{s,-m,n}(r,\theta,\phi)$, $f^{{(1)}\phi}_{s,-m,n}(r,\theta,\phi)$ are $r,\theta, \phi$ components of the spherical wave function respectively.
Again, the current distribution coefficients and the orthogonal basis functions in 3-D space are vectorized by using the index $l'$ ($l'=1,\cdots,3L$) as
\begin{eqnarray}
\label{eq:a}
    a_{l'} &=& \Bigg\{ 
    \begin{array}{ll}
        a_{l'}^{r} & (l'=1,\cdots,L) \\
        a_{l'-L}^{\theta} & (l'=L+1,\cdots,2L) \\
        a_{l'-2L}^{\phi} & (l'=2L+1,\cdots,3L) \\        
    \end{array}, \\
\label{eq:b}
    b_{l'}(u) &=& \Bigg\{ 
    \begin{array}{ll}
        b_{l'}(r) & (l'=1,\cdots,L) \\
        b_{l'-L}(\theta) & (l'=L+1,\cdots,2L) \\
        b_{l'-2L}(\phi) & (l'=2L+1,\cdots,3L) \\        
    \end{array}. 
\end{eqnarray}
Substituting Eqs.\ (\ref{eq:a}) and (\ref{eq:b}) into (\ref{eq:qj}), the SMC is expressed as
\begin{eqnarray}
    q_j &=& \sum_{l'=1}^{3L}
    a_{l'} (-1)^{m+1} \frac{k}{\sqrt[]{\mathstrut \eta}}
    \int_{V_0} f^{{(1)}u}_{s,-m,n}(r,\theta,\phi) b_{l'}(u) \mrm{d}V_0 \nonumber \\
    &=& \sum_{l'=1}^{3L} a_{l'} z_{jl'}, \\
    z_{jl'} &=& (-1)^{m+1} \frac{k}{\sqrt[]{\mathstrut \eta}}
    \int_{V_0} f^{(1)u}_{s,-m,n}(r,\theta,\phi) b_{l'}(u) \mrm{d}V_0,
\end{eqnarray}
where $z_{jl'}$ is calculated from the $j$-th spherical wave function and the $l'$-th orthogonal basis function.
Therefore, the $n$-th antenna's SMC vector $\mbf{q}_n$ is expressed by a matrix $\mbf{Z} \in \mbb{C}^{J \times 3L}$ whose $j$-th column and $l'$-th row component is $z_{jl'}$
and the current distribution coefficient vector $\mbf{a}_n \in \mbb{C}^{3L}$ whose $l'$-th component is $a_{l'}$ as
\begin{eqnarray}
\label{eq:SMC_from_Current2}
    \mbf{q}_n &=& \mbf{Z} \mbf{a}_n.
\end{eqnarray}
From Eq.\ (\ref{eq:SMC_from_Current2}), 
it can be concluded that, if optimal SMC vector $\mbf{q}_n$ is given,
the current distribution coefficient for the optimal antenna directivity can be derived by using a pseudo inverse matrix $\mbf{Z}^{+} \in \mbb{C}^{3L \times J}$ as follows
\begin{eqnarray}
\label{eq:CurrentCoefficient}
	\tilde{\mbf{a}}_n &=& \mbf{Z}^{+} \mbf{q}_n.
\end{eqnarray}
It is noted that the pseudo inverse matrix is calculated by singular values of $\mbf{Z}$ shown in \cite{Golub}.

\section{Numerical analysis}
\label{sec:NumericalAnalysis}
It is noted that the antenna design procedure we proposed is applied for any frequency.
The optimal directivity is derived by using the procedure when the design frequency is determined.
In this section, assuming simple and symmetric angular profiles and sphere antenna volumes defined in wavelength at both sides,
the optimal directivity and the current distribution are derived.

\subsection{Analysis condition} 
In this analysis,  a $2 \times 2$ MIMO system is considered and the analysis condition is shown in Table\ \ref{tbl:AnalysisCondition}. 
We assume that the  angular profile does not depend on the wave number $k$
and define the antenna volume by $kr_0$.
In this case, the analysis results also do not depend on $k$. 

\begin{figure}[tb]
    \centering
    \includegraphics[width=0.45\textwidth]{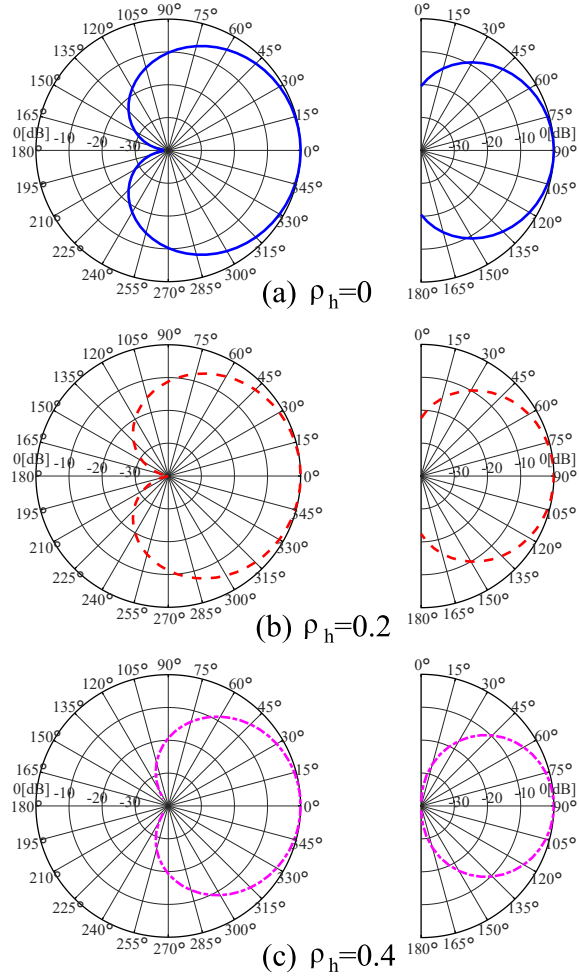}
    \caption{Angular profile with an omni-directional antenna at transmitter or receiver side 
	normalized by the maximum value
	(Left: Horizontal plane, Right: Vertical plane).}
    \label{fig:AngularProfile_cluster}
\end{figure}

The angular profile is strictly defined as a probability density function on the circle such as a von Mises distribution. 
Since the von Mises distribution is approximated by a Gaussian distribution \cite{Mardia}, 
we define the angular profile by using a multivariate Gaussian distribution as follows. 
\begin{eqnarray}
\label{eq:MultivariateGauss}
	P_{\mrm{h}} (\mbf{x}) &\!\!=\!\!& \frac{1}{\sqrt{(2 \pi)^4 \mrm{det}(\mbf{\Sigma})}} 
	\exp \left( -\frac{1}{2} (\mbf{x} - \mbf{m})^\mrm{H} \mbf{\Sigma}^{-1} (\mbf{x} - \mbf{m}) \right) \nonumber \\ \\
	\mbf{x} &\!\!=\!\!& [\theta_\mrm{t}, \phi_\mrm{t}, \theta_\mrm{r}, \phi_\mrm{r}]^\mrm{T} \\
	\mbf{m} &\!\!=\!\!& [\mu_{\mrm{t, \theta}}, \mu_{\mrm{t, \phi}}, \mu_{\mrm{r, \theta}}, \mu_{\mrm{r, \phi}}]^\mrm{T}, 
\end{eqnarray}
where the means of $\theta$ components are $\mu_{\mrm{t, \theta}}, \mu_{\mrm{r, \theta}}$, 
the means of $\phi$ components are $\mu_{\mrm{t, \phi}}, \mu_{\mrm{r, \phi}}$.
A covariance matrix $\Sigma \in \mbb{C}^{4 \times 4}$ is positive definite as shown in \cite{Martin}
and defined as
\begin{eqnarray}
	\mbf{\Sigma} &\!\!=\!\!& (\mbf{c} \mbf{c}^\mrm{H}) 
	\circ \mbf{P} \\
	 \mbf{c} &\!\!=\!\!& [\sigma_{\mrm{t, \theta}}, \sigma_{\mrm{t, \phi}}, \sigma_{\mrm{r, \theta}}, \sigma_{\mrm{r, \phi}}]^\mrm{T} \\	
	\mbf{P} &\!\!=\!\!& \left[ 
	\begin{array}{cccc}
	1 & 0 & \rho_{\mrm{tr},\theta\theta} & \rho_{\mrm{tr},\theta\phi} \\
	0 & 1 & \rho_{\mrm{tr},\phi\theta} & \rho_{\mrm{tr},\phi\phi}\\
	\rho_{\mrm{tr},\theta\theta} & \rho_{\mrm{tr},\phi\theta} & 1 & 0\\
	\rho_{\mrm{tr},\theta\phi} & \rho_{\mrm{tr},\phi\phi} & 0 & 1\\
	\end{array} 
	\right],
\end{eqnarray}
where $\circ$ means Hadmard product, 
the standard deviations of $\theta$ components are $\sigma_{\mrm{t, \theta}}, \sigma_{\mrm{r, \theta}}$, 
the standard deviations of $\phi$ components are $\sigma_{\mrm{t, \phi}}, \sigma_{\mrm{r, \phi}}$.
It is assumed that $\theta_\mrm{t}$ and $\phi_\mrm{t}$, $\theta_\mrm{r}$ and $\phi_\mrm{r}$ are uncorrelated and other components are correlated.
Correlation coefficients between transmitter and receiver sides are defined as $\rho_{\mrm{tr},\theta\theta}, \rho_{\mrm{tr},\theta\phi}, \rho_{\mrm{tr},\phi\theta},\rho_{\mrm{tr},\phi\phi}$.
The propagation environments at the transmitter and receiver are not always independent in the MIMO system due to line-of-sight components, deterministic clusters, and so on,
the correlation between the transmitter and receiver sides should be considered. 
For example, when the correlation coefficient between transmitter and receiver sides is zero, transmitter and receiver sides are independent. 
On the other hand, when the correlation coefficient becomes large, 
the transmitter and receiver sides are not independent and their angular profiles are determined each other.
To simplify the analysis, it is assumed that 
all correlation coefficients have same values 
$\rho_{\mrm{tr},\theta\theta} = \rho_{\mrm{tr},\theta\phi} = \rho_{\mrm{tr},\phi\theta} = \rho_{\mrm{tr},\phi\phi} = \rho_\mrm{h} \geq 0$.
When the correlation coefficients between components of $\mbf{\psi}$ are defined as $\rho_\mrm{h}=0,0.2,0.4$, 
Fig.\ \ref{fig:AngularProfile_cluster} shows the angular profiles using the multivariate Gaussian distributions with an omni-directional antenna at transmitter or receiver side. 
Also, we assume that the angular profile shown in Fig.\ \ref{fig:AngularProfile_cluster} has only $\theta$ polarization component.
Since the $\theta$ polarization component of the angular profile is given, we should optimize only the $\theta$ polarization components of directivities. 
 
\begin{table}[tb]
	\begin{center}
	\caption{Analysis condition.}
	\label{tbl:AnalysisCondition}
		\begin{tabular}{|c|c|}
		\hline
		The number of antennas & $N_{\mrm{t}}=N_{\mrm{r}}=2$ \\
		\hline	
		The angle of departure & $\mu_{\mrm{t, \phi}}=0^{\circ}$, $\mu_{\mrm{t, \theta}}=90^{\circ}$ \\
		\hline
		The angle of arrival & $\mu_{\mrm{r, \phi}}=0^{\circ}$, $\mu_{\mrm{r, \theta}}=90^{\circ}$ \\
		\hline
		The angular spread & $2\sigma_{\mrm{t, \phi}}=60^{\circ}$ \\
		of the transmitter side & $2\sigma_{\mrm{t, \theta}}=30^{\circ}$ \\
		\hline
		The angular spread & $2\sigma_{\mrm{r, \phi}}=60^{\circ}$ \\
		of the receiver side & $2\sigma_{\mrm{r, \theta}}=30^{\circ}$ \\
		\hline
		Polarization of angular profile & Only $\theta$ polarization \\
		\hline
		The radius of the antenna volume & $r_0=\sqrt[]{2}\lambda/4$ \\
		\hline
		The number of modes & $J=16$ \\
		\hline
		The capable difference & 1$\%$ of the difference\\
		\hline
		The number of minute regions &  $L=1600$ \\
		\hline
		\end{tabular}
	\end{center}
\end{table}
\begin{figure}[tb]
	\centering
    \includegraphics[width=0.25\textwidth]{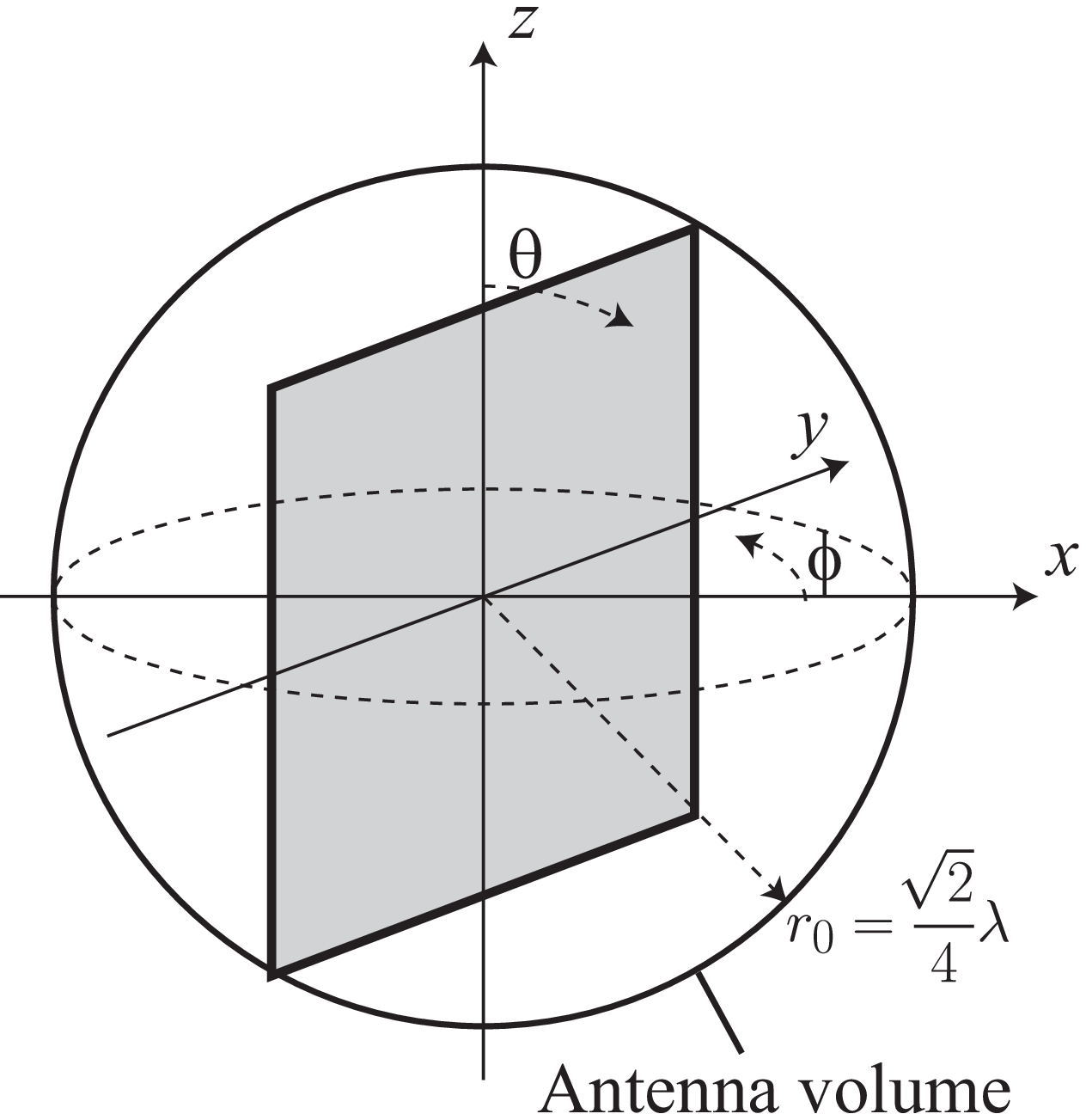}
    \caption{Planar antenna structure.}
    \label{fig:PlanarAntenna}

	\centering
    \includegraphics[width=0.25\textwidth]{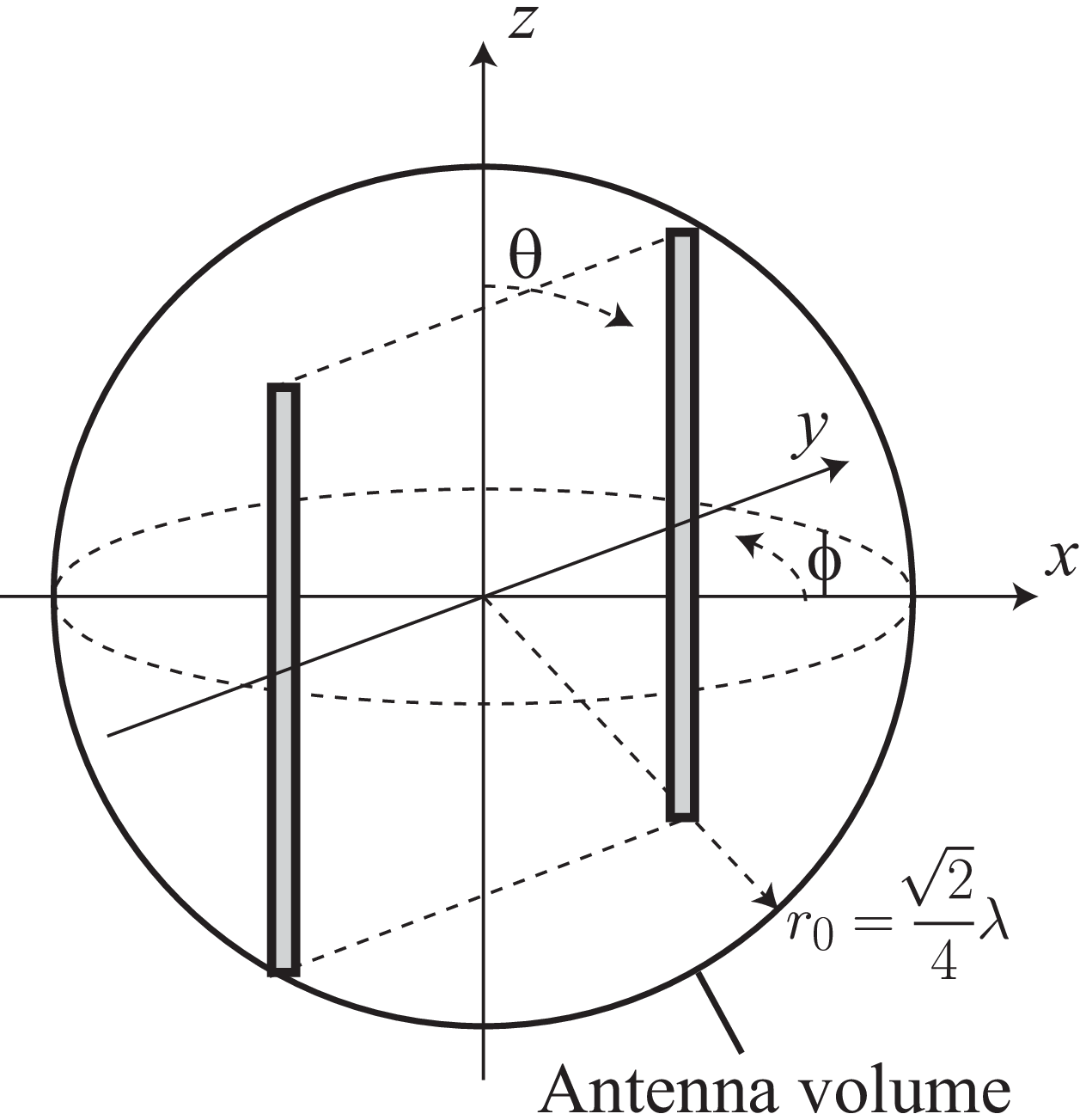}
    \caption{Half-wavelength dipole antenna array.}
    \label{fig:LinearAntenna}
\end{figure}

In this analysis, a planar antenna structure is considered shown in Fig.\ \ref{fig:PlanarAntenna}.
The square of gray color indicates the conductor of the planar antenna 
and two types of current distributions are calculated on the same plane, which correspond to Antenna $\#1$ and Antenna $\#2$ respectively.
A set of conductors including conventional multiple elements is considered as an overall antenna structure
and different patterns of current distribution induced on the structure are called as antennas $\#$1 and $\#$2. 
Since we haven't designed the current distributions element by element, 
the consideration of coupling which conventionally occurs between antenna elements is not needed in this study.
The length of one side of the planar antenna is $\lambda/2$, so the radius of the antenna volume is $r_0=\sqrt{2}\lambda/4$.
Orthogonal basis functions are defined by sine functions, 
which are expressed as 
\begin{eqnarray}
    \vec{b}_l(y,z) &=& b_l(y) \hat{y} + b_l(z) \hat{z}, \\
    b_l(u) &=& \left\{ 
    \begin{array}{ll}
		\frac{\sin\left(k(u-(u_l-\Delta u))\right)}{\sin\left(k\Delta u\right)}
        & (u_l-\Delta u \leq u \leq u_l) \\
		\frac{\sin\left(k((u_l+\Delta u)-u)\right)}{\sin\left(k\Delta u\right)}
        & (u_l\leq u \leq u_l+\Delta u) \\ 
        0 
		& (\mrm{elsewhere}) \\ 
    \end{array}, \right. \\
	u_l &=& \frac{\lambda}{2} \cdot \frac{l}{L}-\Delta u, \\
	\Delta u &=& \frac{\lambda}{4L},
\end{eqnarray} 
where $L$ is the number of minute regions.

Furthermore, we derive the average channel capacity under the same condition in both of the transmitter and receiver 
and compare to the conventional antenna, which is half-wavelength of dipole antenna array shown in Fig.\ \ref{fig:LinearAntenna}.
Conventional multiple planar antennas, like a microstrip antnna array, cannot be located within the spherical volume used in the analysis 
because its antenna array size is larger than the spherical volume. 
Therefore, we choose the half-wavelength dipole antenna array as the conventional antennas because it can be located in the spherical volume. 

The improvement of the capacity is owing to received power gains with two orthogonal directivities.
Thus, we evaluate the received power gains
by calculating the determinant of the channel correlation matrix shown in Eq.\ (\ref{eq:R_cr}).

\subsection{Convergence of objective function}
The maximum average channel capacity is achieved by maximizing the determinant of channel correlation matrix.
as shown in Eq.\ (\ref{eq:Maximize_Det_Rx}).
In our method, 
the transmit or receive antenna directivities are optimized iteratively by using the angular profile with receive or transmit antenna directivities.
Thus, we should confirm the convergence of the objective function i.e. the determinant of the channel correlation matrix. 
Not only the increase of antenna gain 
but also the orthogonality of the antenna directivity weighted by the angular profile 
affects the increase of the objective function
 and results in improvement of channel capacity. 
Therefore, both the antenna gain and the orthogonality of the antenna directivity weighted by the angular profile should be considered at the same time. 
When the antenna gain increases, 
the diagonal components of the channel correlation matrix become large. 
And when the orthogonality of the antenna directivity weighted by the angular profile increases, 
the non-diagonal components of the channel correlation matrix decrease. 
Because these two factors increase the value of the determinant of correlation matrix, we compared the values of determinant in the next section. 

\begin{figure}[tb]
    \centering
    \includegraphics[width=0.4\textwidth]{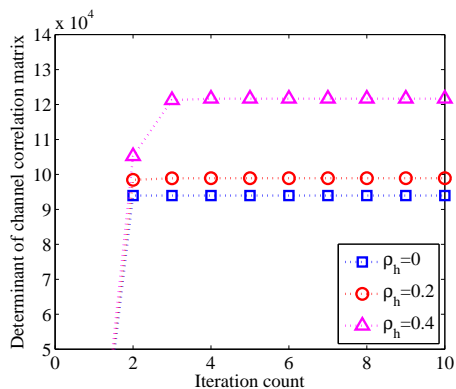}
    \caption{Objective function of Eq. (\ref{eq:Maximize_Det_Rx}).}
    \label{fig:ObjectiveFunc}
\end{figure}

Figure\ \ref{fig:ObjectiveFunc} shows the determinant of the channel correlation matrix calculating iteratively.
In the analysis, the initial SMCs are defined as those of the half-wavelength dipole antenna array.
The capable difference $\epsilon$ is defined as 1$\%$ of the difference of the determinant in the previous calculation, 
which is $0.01 \left|\det \bar{\mbf{R}}_{\mrm{c,t}}^{(itr-1)} - \det \bar{\mbf{R}}_{\mrm{c,r}}^{(itr-2)} \right|$ 
or  $0.01 \left|\det \bar{\mbf{R}}_{\mrm{c,r}}^{(itr-1)} - \det \bar{\mbf{R}}_{\mrm{c,t}}^{(itr-2)} \right|$ 
for the $itr$-th times calculation.
It is because the directivities are converged enough in the iterative calculation when the difference is within 1 \% in the analysis.
The determinant is normalized by that with the half-wavelength dipole antenna array shown in Fig.\ \ref{fig:LinearAntenna}. 
When the count is zero, the half-wavelength dipole antenna array is used at the transmitter and receiver.
The odd count means the optimization of receive antennas 
and the even count means the optimization of transmit antennas.
From the results, we confirmed that 
the determinant of the channel correlation matrix converges in the case of several correlation coefficients.
From now, we shall use the case of $\rho_\mrm{h}=0.2$ for analyses.

\subsection{Optimal antenna directivity and current distribution}
\label{sec:NumericalAnalysis_Directivity}
\begin{figure}[tb]
    \centering
    \includegraphics[width=0.48\textwidth]{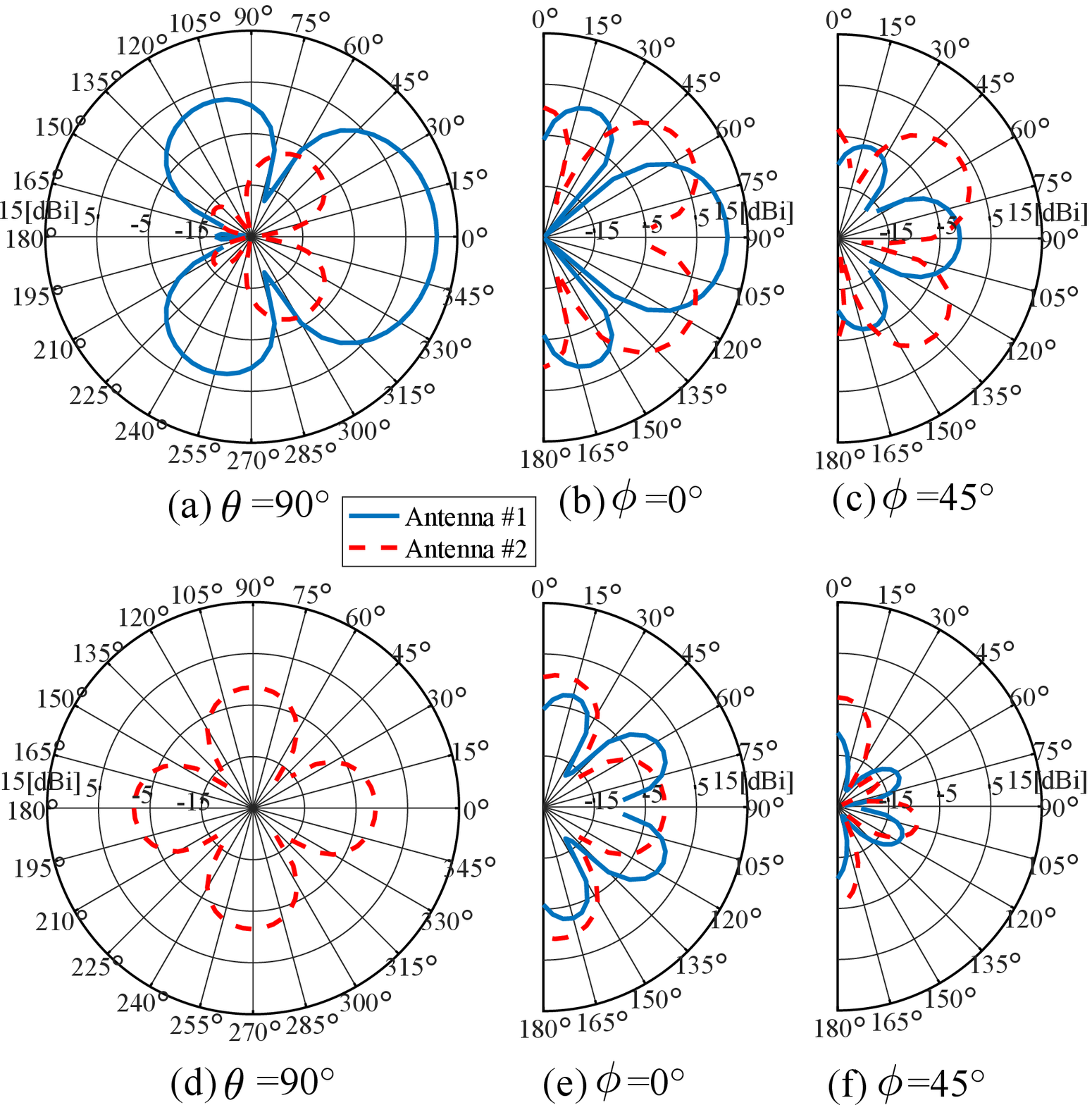}
    \caption{Optimal antenna directivities ((a)-(c): $\theta$ polarization component, (d)-(f): $\phi$ polarization component).}
    \label{fig:OptimalDirectivity_vpol}

    \centering
    \includegraphics[width=0.4\textwidth]{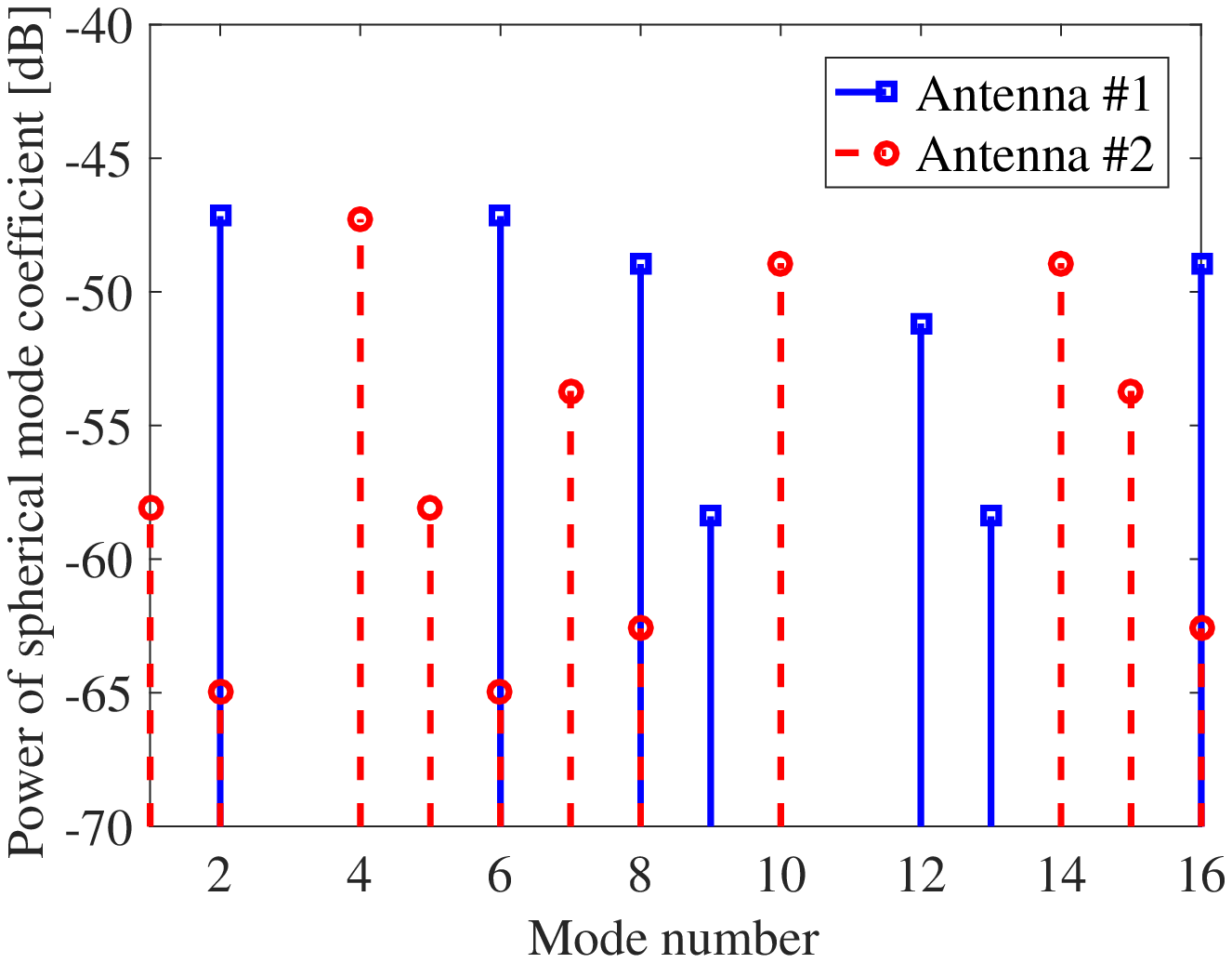}
    \caption{Spherical mode coefficients of optimal antenna directivities.}
    \label{fig:OptimalSMC_vpol}
\end{figure}

The same directivities are derived at the transmitter and receiver 
because the angular profile is defined symmetry at the transmitter and receiver, 
thus we show the results at the receiver.
Optimal directivities and optimal SMCs are derived in Figs.\ \ref{fig:OptimalDirectivity_vpol} and \ref{fig:OptimalSMC_vpol}.
The directivities are shown in the case of a horizontal plane ($\theta\!=\!90^\circ$) 
and vertical planes ($\phi\!=\!0^\circ, 45^\circ$).
It is found that the directivity of Antenna $\#$1 has a peak and that of Antenna $\#$2 has a null toward direction 
from which waves with high intensity come in $\theta$ polarization.
The directivity of Antenna $\#$2 has a peak toward $\theta\!=\!60^\circ$ and $\phi\!=\!45^\circ$.
In this case, the determinant of channel correlation matrix is 50 dB larger than that with the half-wavelength dipole antenna array.

\begin{figure}[tb]
    \centering
    \includegraphics[width=0.45\textwidth]{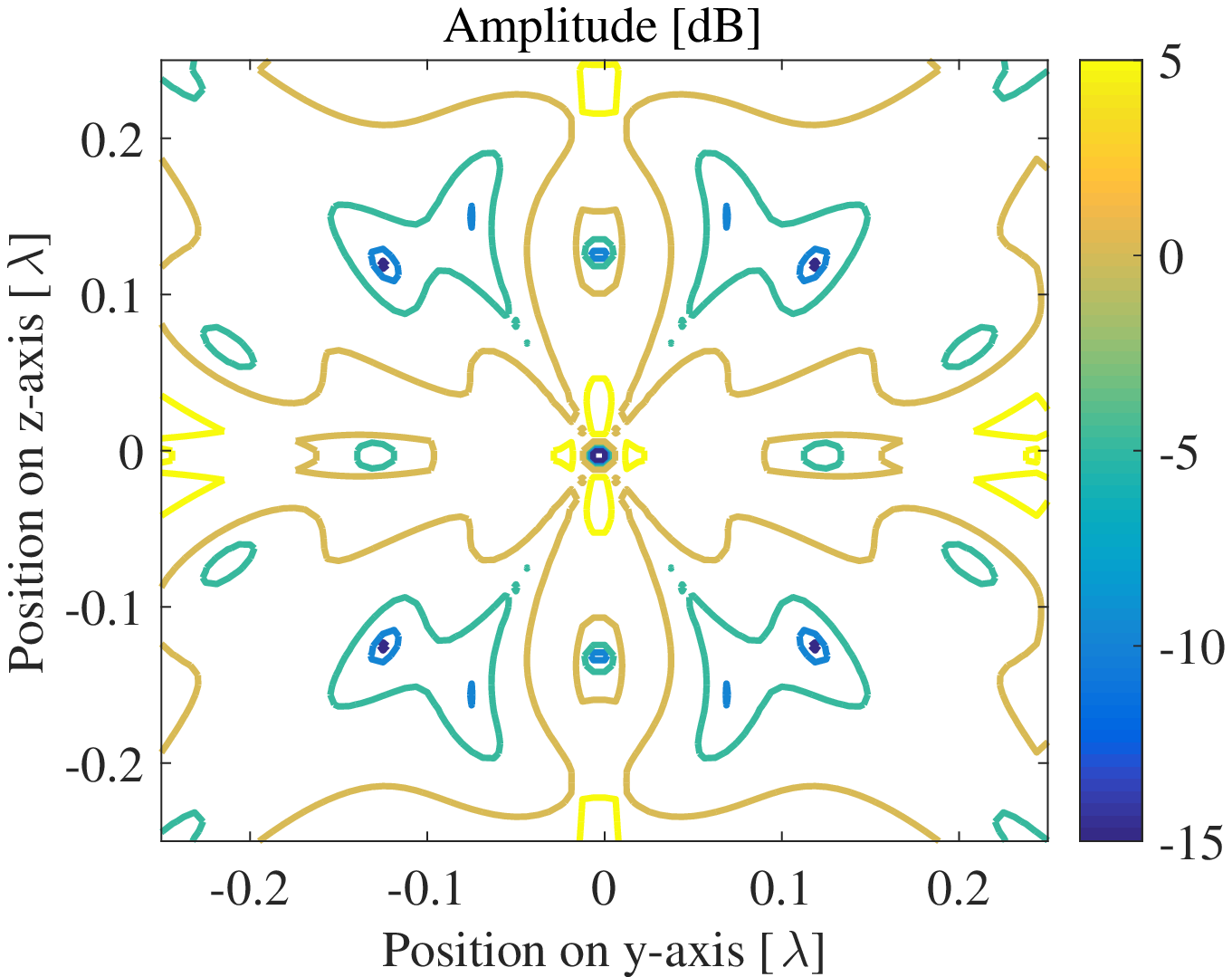}
    \caption{Current distribution for antenna $\# 1$ (Amplitude).}
    \label{fig:Current1_A}

    \centering
    \includegraphics[width=0.45\textwidth]{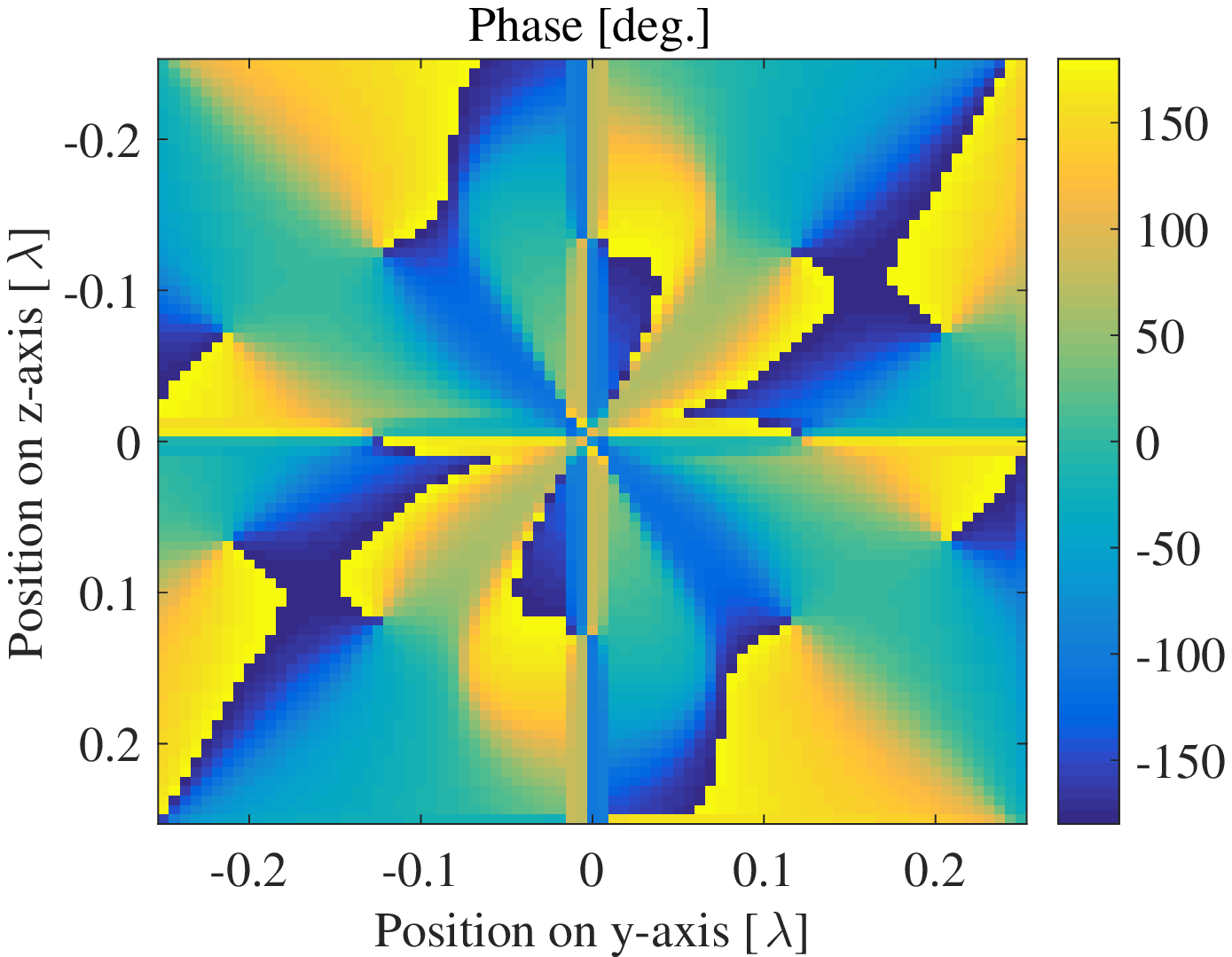}
    \caption{Current distribution for antenna $\# 1$ (Phase).}
    \label{fig:Current1_P}
\end{figure}

\begin{figure}[tb]
    \centering
    \includegraphics[width=0.45\textwidth]{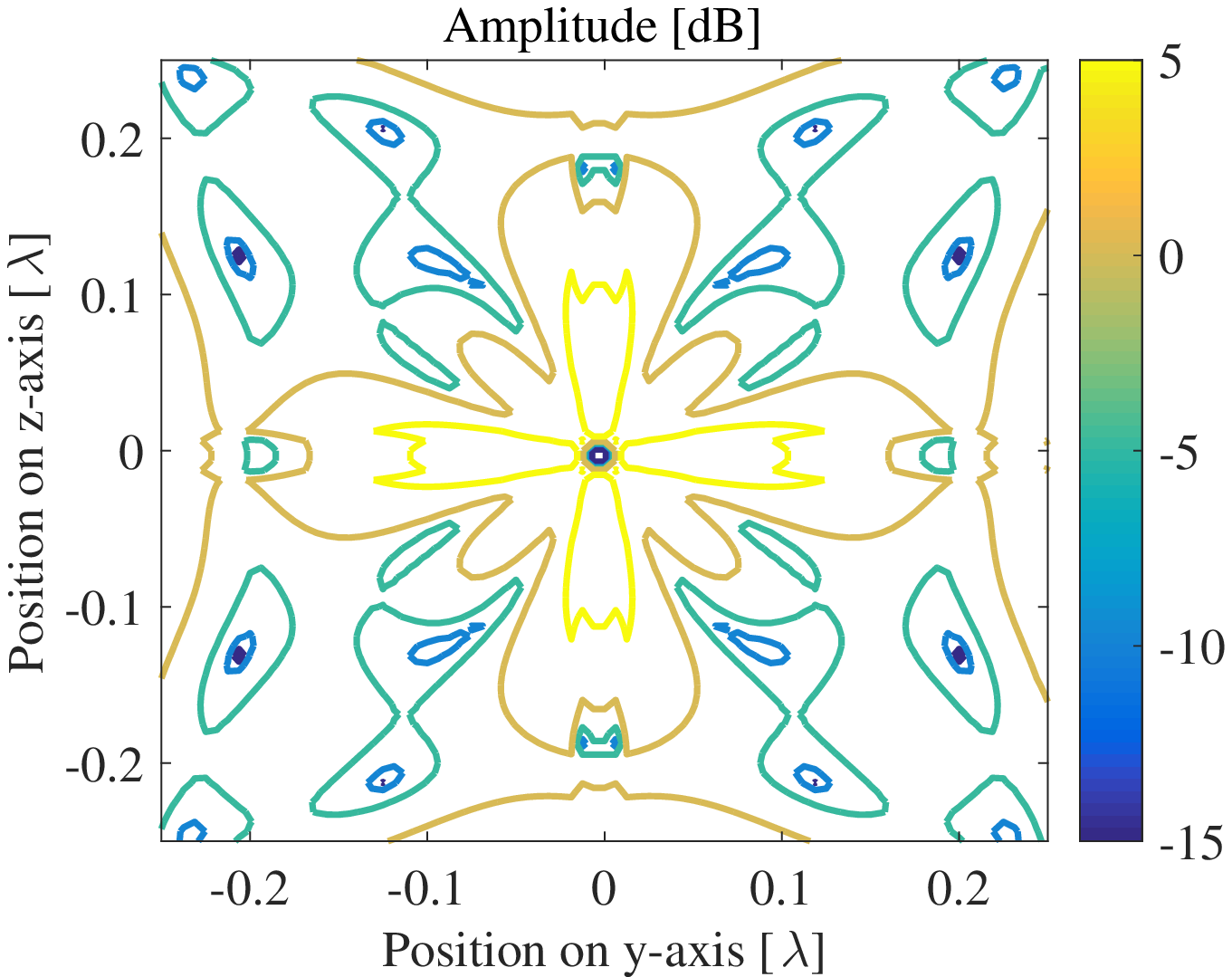}
    \caption{Current distribution for antenna $\# 2$ (Amplitude).}
    \label{fig:Current2_A}

    \centering
    \includegraphics[width=0.45\textwidth]{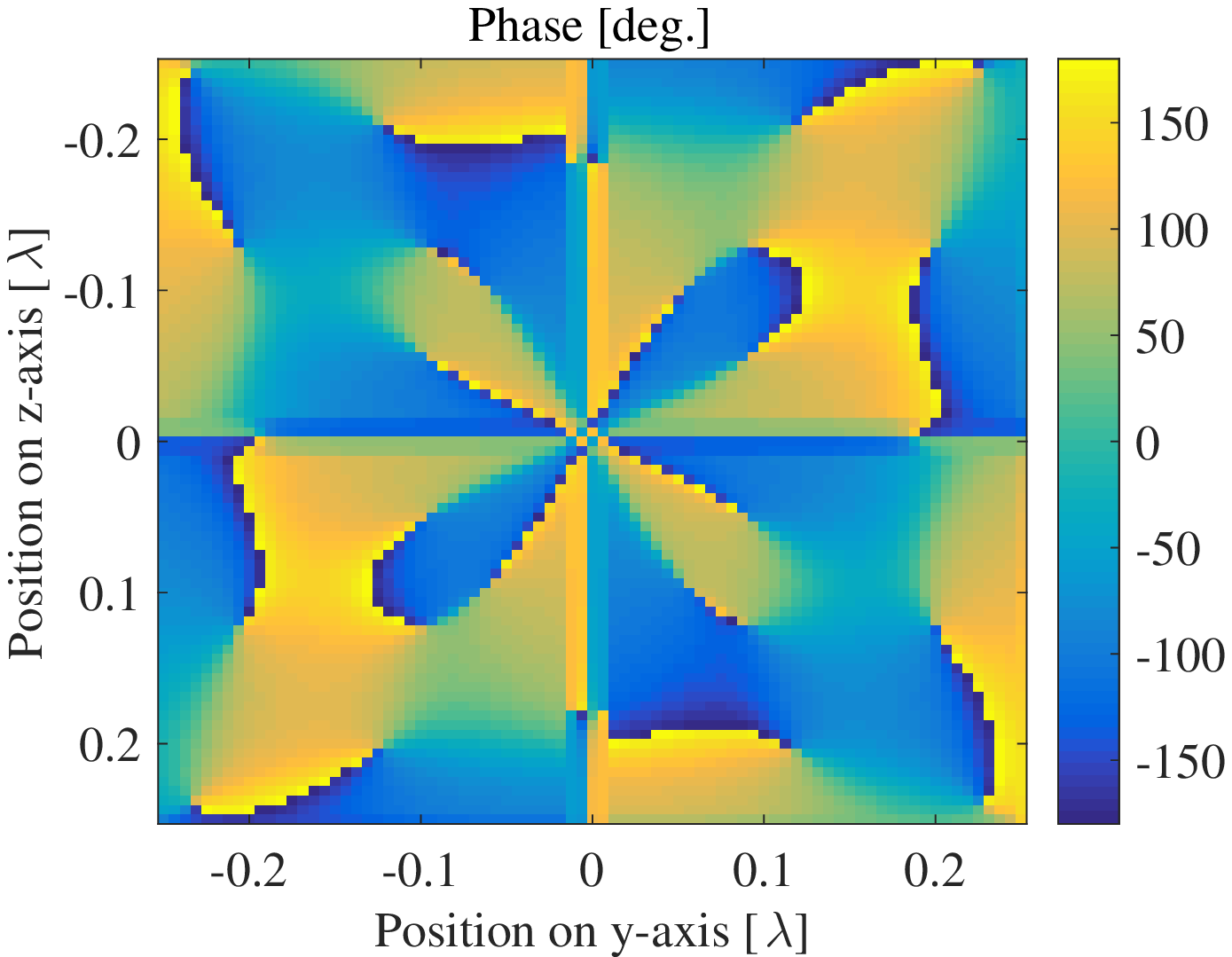}
    \caption{Current distribution for antenna $\# 2$ (Phase).}
    \label{fig:Current2_P}
\end{figure}

The current distributions for planar antennas are calculated and shown in 
Figs.\ \ref{fig:Current1_A}-\ref{fig:Current2_P}.
Each contour line indicates the amplitude of current distribution
and each color indicates the phase of current distribution on the planar antenna structure of Fig.\ \ref{fig:PlanarAntenna}.
One method to realize the derived current distributions is to divide the surface into many small regions 
and to feed them with corresponding excitation coefficients by using spatial sampling theory of the current distribution. 
However, it's obviously not efficient in terms of hardware cost. 
Therefore, we'd like to keep it as a future work and derive novel feeding structure with limited costs.

\begin{figure}[tb]
    \centering
    \includegraphics[width=0.48\textwidth]{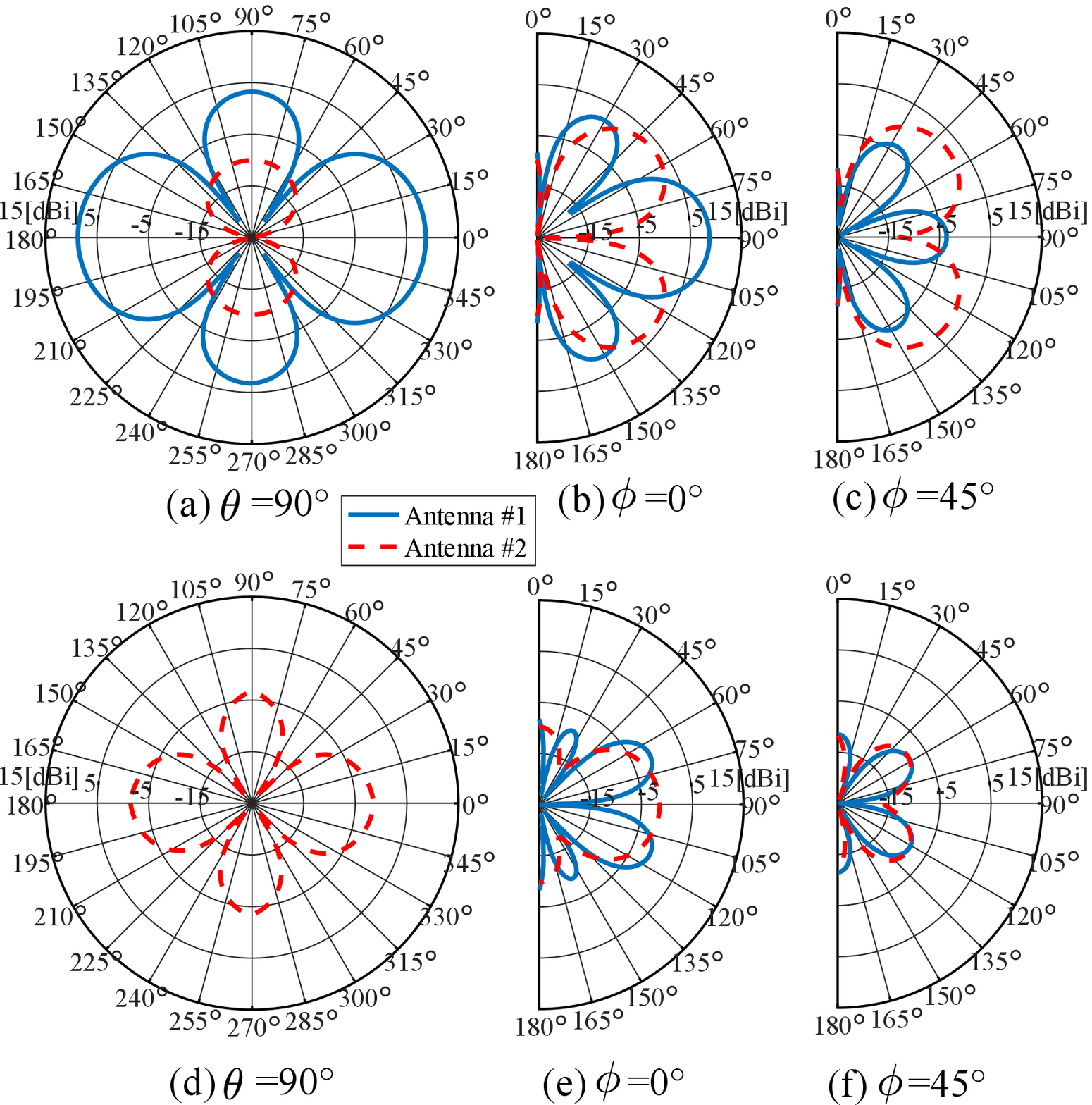}
    \caption{Recalculated planar antenna directivities  ((a)-(c): $\theta$ polarization component, (d)-(f): $\phi$ polarization component).}
    \label{fig:Directivity2D_vpol}

    \centering
    \includegraphics[width=0.4\textwidth]{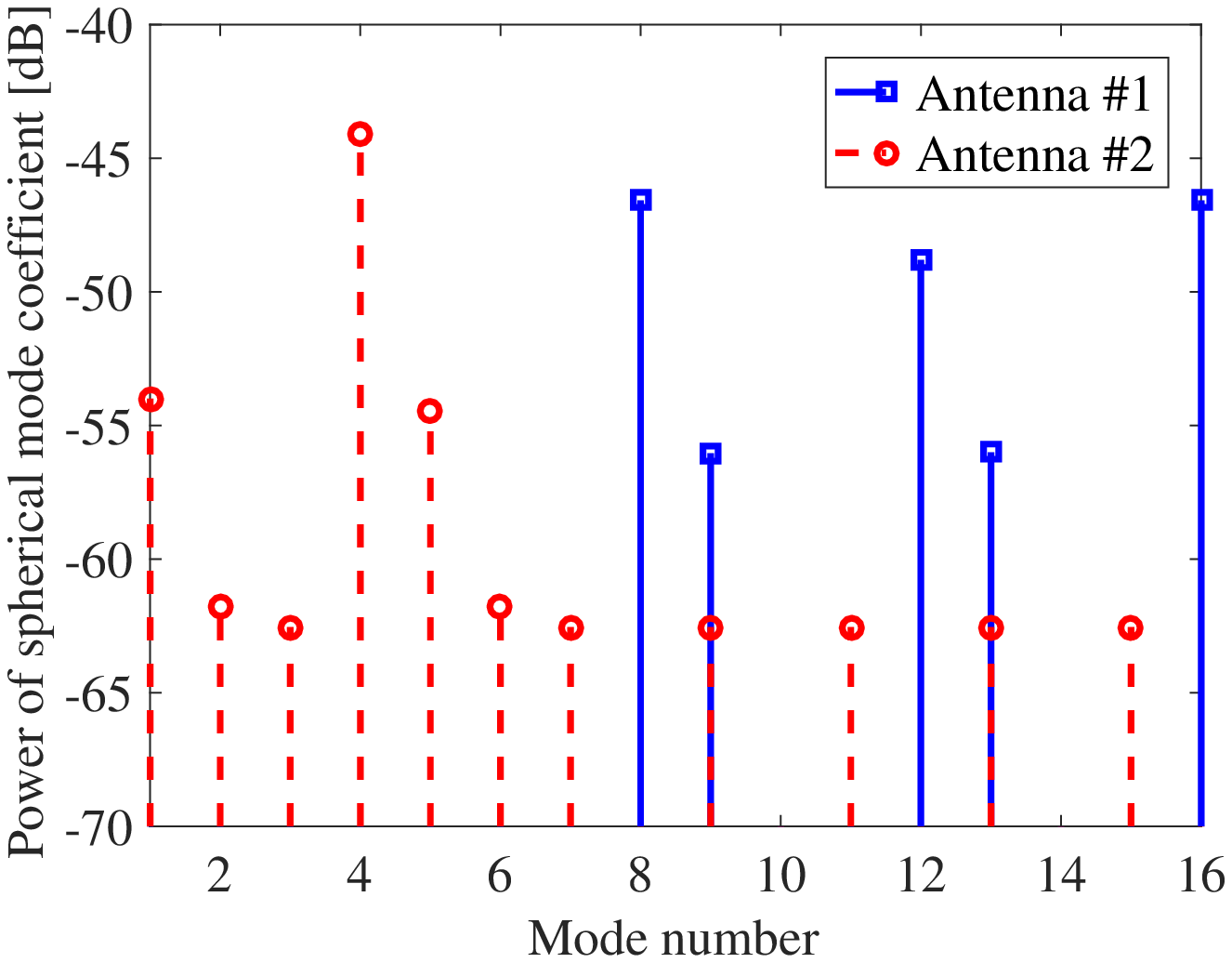}
    \caption{Spherical mode coefficients of recalculated antenna directivities.}
    \label{fig:SMC2D_vpol}
\end{figure}

The directivity and the SMCs recalculated from each current distribution are shown in Figs.\ \ref{fig:Directivity2D_vpol} and \ref{fig:SMC2D_vpol}.
The peak and null is almost the same with that of the optimal directivity in the vertical plane.
However, the directivities in horizontal plane are different because of a symmetry of the planar antennas' structures.
The optimal directivity in Fig.\ \ref{fig:OptimalDirectivity_vpol} is derived 
without any limitation of the antenna configuration. 
On the other hand, the directivity in Fig.\ \ref{fig:Directivity2D_vpol} is derived 
with the constraint of planar structure without thickness on $yz$-plane 
as in Fig.\ \ref{fig:PlanarAntenna}. 
Since this structure has a symmetrical property with respect to the $yz$-plane, 
the directivity must have the symmetrical pattern and therefore the gain for $\phi\!=\!0^\circ$ is decreased. 

It is also found that the effective modes of optimal directivities and recalculated planar antennas are the same.
The determinant of the channel correlation matrix achieves 42 dB larger than that with the half-wavelength dipole antenna array.

\subsection{Average channel capacity}
\begin{figure}[tb]
    \centering
    \includegraphics[width=0.42\textwidth]{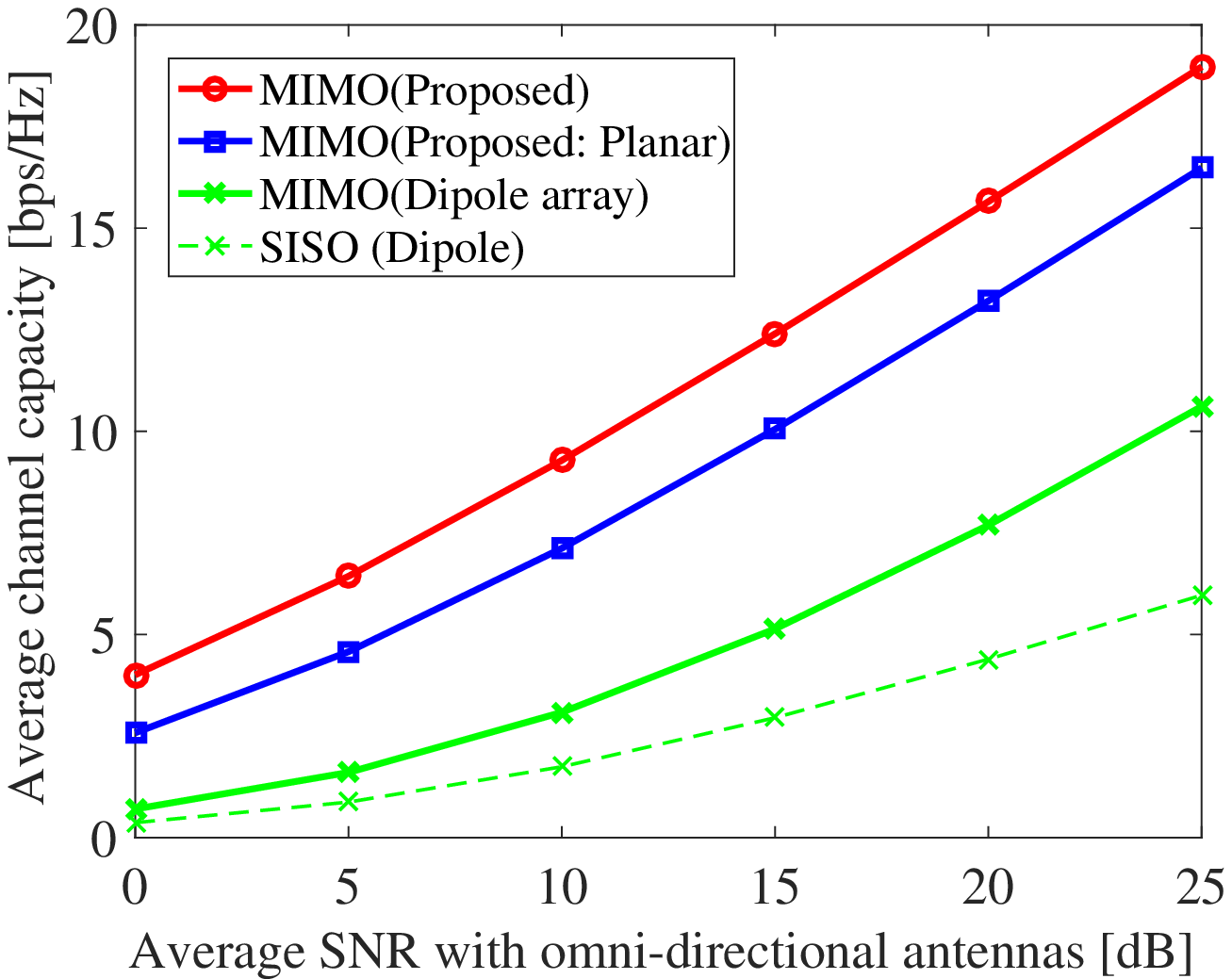}
    \caption{Comparison of average channel capacity.}
    \label{fig:Capacity_vpol}
\end{figure}

Figure\ \ref{fig:Capacity_vpol} indicates the average channel capacity in four cases as follows. 
\begin{itemize}
\item Optimal Directivity: Proposed.
\item Recalculated Planar antenna directivity: Proposed (Planar).
\item Half-wavelength dipole antenna array: Dipole array.
\item Single-Input Single-Output with Half-wavelength dipole antenna: SISO.
\end{itemize}
When the average signal-to-noise ratio (SNR) is 15 dB, 
the channel capacity with the optimal directivity that we proposed is
7.3 bps/Hz larger than that with Dipole array 
and 9.4 bps/Hz larger than that of SISO.
The recalculated planar antenna directivity is degraded 2.3 bps/Hz compared to that with the optimal directivity.
From the results shown in subsection \ref{sec:NumericalAnalysis_Directivity}, 
the determinant of the channel correlation matrix with the recalculated directivity is much larger than that with the dipole array 
because the directivity is matched to the angular profile 
and the loss of radiation power is minimized.
Therefore, large channel capacity is achieved by the directivities matched to the angular profile.

\section{Conclusion}
\label{sec:Conclusion}
We proposed a method to derive the optimal directivity and the current distribution by using spherical mode expansion
in order to maximize the average channel capacity.
Maximizing the capacity is equivalent to maximizing the determinant of the channel correlation matrix, 
thus the SMC vector can be derived from the eigen vector of the spherical mode correlation matrix. 
Using the proposed method, the orthogonal directivities are derived 
and lower correlation is achieved when the angular profile and the antenna volume are given.
The SMCs can also be derived from the current distribution and spherical wave functions of $c=1$.
Consequently, the current distribution coefficients for the optimal directivity are derived 
by using a pseudo inverse matrix of $\mbf{Z}$ calculated from spherical wave functions of $c=1$ and orthogonal basis functions.

In numerical analysis, 
we derived an example of the optimal directivity and recalculated planar antenna directivity.
From the analysis results, 
we confirmed that the directivity optimization that we proposed improves the average channel capacity of MIMO systems.

Since the main contribution of this paper is the derivation of optimal directivities and corresponding current distributions, we would like to remain the design of feeding points as our future work.
To determine the feeding points is necessary for actual antenna design, 
thus we would like to derive novel feeding structure with limited costs in our future work.
for example we will apply a theory about matrix of decoupling and matching networks \cite{Kagoshima} 
or an expanded design manner for massive MIMO antennas.
Also, in a future study, 
we should figure out how to obtain joint angular profile when channels are varying.


{\bf Maki Arai} received the B.E. and M.E. degrees in electrical and electronic engineering
from the Tokyo Institute of Technology, in 2010 and 2012. 
She joined the NTT Network Innovation Laboratories, Nippon Telegraph and Telephone Corporation (NTT) in 2012.
Her current research interests are high speed wireless communication systems and analysis and design of MIMO antennas.
She received the Research Encouragement Award of the Institute of Electrical Engineers of Japan (IEEJ) in 2010, 
the Antenna and Propagation Research Commission Student Award from the Institute of Electronics, Information and Communication Engineers (IEICE) in 2012, 
and the Young Engineers Award from the IEICE in 2015. 
She is a member of IEEE and IEICE.

{\bf Masashi Iwabuchi} received the B.E. and M.E. degrees 
from Tokyo Institute of Technology, Tokyo, Japan, in 2008 and 2010, respectively. 
In 2010, he joined NTT Access Network Service Systems Laboratories, 
Nippon Telegraph and Telephone Corporation (NTT), in Japan. 
He is engaged in the research of medium access protocol and resource management 
for next generation wireless communications. 
He received the young investigators award from IEICE in 2015.

{\bf Kei Sakaguchi} received the B.E. degree in electrical and computer
engineering from Nagoya Institute of Technology, Japan in 1996, and
the M.E. degree in information processing from Tokyo Institute of
Technology, Japan in 1998, and the Ph.D. degree in electrical and
electronic engineering from Tokyo Institute of technology in 2006.
From 2000 to 2007, he was an Assistant Professor at Tokyo Institute of Technology.
Since 2007, he has been an Associate Professor at the same university.
Since 2012, he has also joined in Osaka University as an Associate Professor.
He received the Young Engineering Awards from IEICE and
IEEE AP-S Japan Chapter in 2001 and 2002 respectively, 
the Outstanding Paper Awards from SDR Forum and IEICE in 2004 and 2005, respectively, 
the Tutorial Paper Award from IEICE Communication Society in 2006, 
and the Bast Paper Awards from IEICE Communication Society in 2012, 2013, and 2015. 
He is currently playing roles of the General Chair in IEEE WDN-CN2015, 
the Honorary General Chair in IEEE CSCN2015, 
and the TPC Chair in IEEE RFID-TA2015.
His current research interests are 5G cellular networks, sensor networks, and wireless energy transmission. 
He is a member of IEEE.

{\bf Kiyomichi Araki} received the Ph.D. degree 
from Tokyo Institute of Technology, Japan, in 1978. 
In In 1979-1980 and 1993-1994, he was a visiting research scholar at University of Texas, 
Austin and University of Illinois, Urbana, respectively. 
Since 1995 to 2014 he has been a Professor at Tokyo Institute of Technology, 
and now an Emeritus Professor. 
He has numerous journals and peer review publications in RF ferrite devices, 
RF circuit theory, electromagnetic field analysis, software defined radio, array signal processing, 
UWB technologies, wireless channel modeling, MIMO communication theory, 
digital RF circuit design, information security, and coding theory.

\end{document}